\documentclass[]{aa}  
\usepackage{graphicx}
\usepackage{txfonts}
\usepackage[colorlinks,citecolor=blue,urlcolor=blue]{hyperref}
\usepackage{verbatim}
\usepackage{float}
\bibpunct{(}{)}{;}{a}{}{,}

\begin{document} 

\title{Whispering in the dark}
\titlerunning{Whispering in the dark - faint X-ray emission from black holes with OB star companions}

  \subtitle{Faint X-ray emission from black holes with OB star companions}

\author{K. Sen \inst{1}\thanks{The first two authors have contributed equally to this work}
    \and
    I. E. Mellah\inst{2,3}\footnotemark[1]
    \and 
    N. Langer\inst{4,5}
    \and
    X.-T. Xu\inst{4}
    \and
    M. Quast\inst{6}
    \and
    D. Pauli\inst{7}
    }

\institute{
    Institute of Astronomy, Faculty of Physics, Astronomy and Informatics, Nicolaus Copernicus University, Grudziadzka 5, 87-100 Torun, Poland \\
    \email{ksen@umk.pl}
    \and
    Departamento de Física, Universidad de Santiago de Chile, Av. Victor Jara 3659, Santiago, Chile
    \and 
    Center for Interdisciplinary Research in Astrophysics and Space Exploration (CIRAS), USACH, Santiago, Chile
    \and
    Argelander-Institut f\"ur Astronomie, Universität Bonn, Auf dem H\"ugel 71, 53121 Bonn, Germany
    \and
    Max-Planck-Institut f\"ur Radioastronomie, Auf dem H\"ugel 69, 53121 Bonn, Germany
    \and
    Institut f\"ur Physik, Otto-von-Guericke Universit\"at, Universit\"atsplatz 2, 39106 Magdeburg, Germany
    \and
    Institut f\"ur Physik und Astronomie, Universit\"at Potsdam, Karl-Liebknecht-Str. 24/25, 14476 Potsdam, Germany
    }

\date{Received May 31, 2024; accepted July 27, 2024}
 
\abstract
  {Recently, astrometric and spectroscopic surveys of OB stars have revealed a few stellar-mass black holes (BHs) with orbital periods as low as 10 days. Contrary to wind-fed BH high-mass X-ray binaries, no X-ray counterpart has been detected, probably because of the absence of a radiatively efficient accretion disk around the BH. Yet, dissipative processes in the hot, dilute and strongly magnetized plasma around the BH (so-called BH corona) can still lead to non-thermal X-ray emission (e.g. synchrotron).}
  {We determine the X-ray luminosity distribution from BH+OB star binaries up to orbital periods of a few thousand days.}
  {We use detailed binary evolution models computed with MESA for initial primary masses of 10-90\,$M_{\odot}$ and orbital periods from 1-3000\,d. The X-ray luminosity is computed for a broad range of radiative efficiencies that depend on the mass accretion rate and flow geometry.} 
  {For typical conditions around stellar-mass BHs, we show that particle acceleration through magnetic reconnection can heat the BH corona. A substantial fraction of the gravitational potential energy from the accreted plasma is converted into non-thermal X-ray emission. Our population synthesis analysis predicts at least 28 (up to 72) BH+OB star binaries in the Large Magellanic Cloud (LMC) to produce X-ray luminosity above 10$^{31}$\,erg$\,$s$^{-1}$, observable through focused Chandra observations. We identify a population of SB1 systems in the LMC and HD96670 in the Milky Way comprising O stars with unseen companions of masses above 2.3\,$M_{\odot}$ that aligns well with our predictions and may be interesting sources for follow-up observations. The predicted luminosities of the OB companions to these X-ray-emitting BHs are 10$^{4.5-5.5}$\,$L_{\odot}$.}
  {These results make the case for long-time exposure in X-rays of the stellar-mass BH candidates identified around OB stars. It will constrain the underlying population of X-ray-faint BHs, the evolution from single to double degenerate binaries, and the progenitors of gravitational wave mergers.}


   \maketitle
%

\section{Introduction}

The detection of merging stellar-mass black holes (BHs) and neutron 
stars \citep{abbott2016,abbott2019} has ushered in a thrilling quest 
to discover their progenitors. Co-evolution of massive binary stars 
(i.e. hydrogen burning starts at the same instant of time in both 
binary components) is a possible formation channel of merging compact 
objects \citep{Belczynski2008,Marchant2016,kruckow2018,Spera2019,vigna-gomez2019,Mapelli2020,Bavera2021,Marchant2021,Jiang2023}. The stable 
mass transfer channel has been shown to significantly contribute to 
double compact object mergers \citep{Heuvel2017,Gallegos-Garcia2021,vanSon2022,Shao2022,Briel2023,Picco2024,Dorozsmai2024,Olejak2024}. 
However, many unconstrained assumptions (such as mass and angular 
momentum loss, internal mixing, stellar winds) used in the modelling 
of binary evolution cripple our capacity to make reliable predictions 
regarding the rates and properties of double compact object mergers 
\citep{Broekgaarden2022,Mandel2022,Marchant2024}. 

In the co-evolutionary scenario, high-mass X-ray binaries (HMXBs) 
represent an ephemeral albeit decisive phase before the formation 
of double compact objects from massive binaries. HMXBs contain a 
massive star in orbit with an X-ray bright compact object 
\citep{walter2015,Kretschmar2019,Motta2021,Chaty2022,Fortin2023}. 
A few of them were found to host stellar-mass BHs accreting stellar 
material from mass transfer via Roche lobe overflow and/or the 
capture of a fraction of the stellar wind (e.g. Cygnus X-1 
\citealp{Orosz2011}, and LMC X-1 \citealp{Orosz2009}). Beyond 
their role in explaining mergers, these systems are excellent 
testbeds for theories of stellar evolution \citep{Belczynski2008,Fishbach2022}, 
gravity and magnetism in the strong field regime 
\citep{Narayan1995,Bozzo2008,Karino2019}, stellar winds and 
X-ray irradiation \citep{Krticka2018,Vilhu2021}. 

The plasma surrounding accreting stellar-mass BHs is a source of 
X-rays whose intensity and spectral properties depend on the flow 
geometry (disk or sphere-like) and on the mass accretion rate 
\citep{Shakura1973,Narayan1995,Frank2002}. At high mass accretion 
rates, a geometrically thin and optically thick accretion disk can 
form which is radiatively efficient and X-ray bright. But at lower 
mass accretion rates \citep{Sharma2007,Xie2012} and/or when the 
flow is not centrifugally supported \citep{Narayan1994}, an 
accretion disk seldom forms and the radiative efficiency of the 
plasma drops. In very sub-Eddington accretion flows, most of the 
kinetic energy of the electrons is not radiated away but advected 
into the BH event horizon \citep[ADAF,][]{Ichimaru1977,Narayan1994,Narayan1998,Yuan2001}. 
Also, a significant fraction of the material captured by the 
gravitational potential of the BH might eventually not accrete 
onto the BH but is lost in strong winds (ADIOS -- \citealp[]{Blandford1999,Yuan2012b,Yuan2015}). 
As a result, the X-ray luminosity of most stellar-mass BHs might 
be below the detectability threshold of all-sky instruments 
($\sim$\,$10^{35}$\,erg$\,$s$^{-1}$, taken from \citealp[]{vanbeveren2020}), 
although they can be detected during outbursts. 

Population synthesis models predict thousands of BHs in orbit with 
a main sequence, giant or supergiant companions in the Milky Way 
\citep{Shao2019,Shao2020}. Yet, most BHs in a binary with a main 
sequence OB star are expected to be X-ray faint due to the inability 
to form a radiatively efficient accretion disk around the BH 
\citep{Sen2021,Hirai2021}. In the absence of an accretion disk, 
X-rays are dominated by non-thermal emission from electrons 
accelerated in the hot, dilute, magnetized and spherical (i.e. 
non-centrifugally maintained) region around the BH, called the 
BH ``corona" \citep{Bisnovatyi1997,Quataert1999}. This radiative 
model has been corroborated by observations of supermassive BHs 
like Sagittarius A$^*$ \citep{Baganoff2003}, M87$^*$ \citep{EHT2019} 
and NGC\,3115 \citep{Wong2011}. In this work, we use the above model 
to investigate the properties of X-ray emission from stellar-mass 
BHs in orbit with a main sequence OB star.

The sample of detected stellar-mass BHs is likely very incomplete. 
The Milky Way and Magellanic Clouds are predicted to harbour millions 
and hundreds of stellar-mass BHs respectively 
\citep{Timmes1996,Breivik2017,Langer2020,Chawla2022,Jayasinghe2023}. 
Yet, only a few tens were confirmed through X-ray detection 
\citep{walter2015,Corral2016} and, more recently, a handful through 
monitoring of the stellar companion 
\citep{Giesers2018,Thompson2019,Liu2019,Rivinius2020,Masuda2021,Jayasinghe2021,Gomez2021,Saracino2022,Shenar2022n,Tanikawa2023,El-Badry2023a,El-Badry2023b,Chakrabarti2023,Gaia2024}. 
Most of the $\sim$\,60 stellar-mass BHs detected through X-ray emission 
are transients, with periodicities of the order of years to decades 
\citep{McClintock2006b,Corral2016}. In contrast, their X-ray luminosity 
during quiescence is several orders of magnitude lower than 
$10^{35}$erg$\,$s$^{-1}$.

The Large Magellanic Cloud (LMC) represents an ideal laboratory to 
hunt for stellar mass BHs with the above techniques owing to its 
homogeneous sample of well-characterised massive stars and low 
interstellar extinction compared to the Milky Way 
\citep{Evans2011,sana2013,Evans2015,Villasenor2021,mahy2019a,Shenar2022}. 
Several single-lined spectroscopic (SB1) binaries have been identified 
\citep{Villasenor2021,Shenar2022}. The \textit{Chandra} T-ReX 
programme \citep{Crowther2022} has shown that faint X-ray 
sources down to 10$^{31}$\,erg$\,$s$^{-1}$ can be detected, 
but the statistics of X-ray-faint (10$^{31}$-10$^{35}$\,erg$\,$s$^{-1}$) 
BHs is still unclear. We do a population synthesis study to predict the 
X-ray luminosity function of BH+OB binaries in the LMC with specific 
attention to the faint end of the distribution. We discuss the possibility 
of confirming the presence of a candidate BH through targeted X-ray 
searches with \textit{Chandra} 
\citep{Evans2018}, and SRG/eROSITA \citep{Predehl2021,Sunyaev2021}. 

Section\,\ref{section_radiative_efficiency} investigates the radiative 
mechanisms around BHs accreting at low mass accretion rates. Section\,\ref{section_method} 
describes the detailed binary evolution models and our assumptions 
to estimate the X-ray luminosity during the BH+OB star phase from 
the models. We describe our results in Sect.\,\ref{section_results} 
and compare our model predictions with observations in Sect.\,\ref{section_observations}. 
We discuss relevant uncertainties and observational challenges 
in Sect.\,\ref{section_discussion}. Section\,\ref{section_conclusion} 
summarises the takeaway messages from our work and future directions. 


\section{Radiative efficiency}
\label{section_radiative_efficiency}

When a black hole accretes plasma at a rate $\dot{M}_{\rm net}$ (in 
$M_{\odot}\,$yr$^{-1}$, measured within a few times the event horizon's 
radius), the release of gravitational potential energy is susceptible 
to produce X-rays with a luminosity $L_X=\epsilon \dot{M}_{\rm net}c^2$, 
with $\epsilon$ the radiative efficiency and $c$ is the speed of light 
in vacuum. The radiative efficiency of this plasma (and, in turn, the 
spectral properties of the emission) depends on the flow geometry and 
the mass accretion rate \citep[][]{Liu2022}. If the plasma is supplied 
by a Roche lobe overflowing stellar companion, a disk unavoidably forms 
\citep{Savonije1978}. If the companion to the BH is not Roche-lobe 
filling, the flow geometry is set by the specific angular momentum $l$ 
of the accreted plasma from the stellar wind of the companion 
(Sect.\,\ref{sec:disk_formation}). 

The accretion regime is determined by the ratio 
$\dot{M}_{\rm net}/\dot{M}_\mathrm{Edd}$, where $\dot{M}_\mathrm{Edd}$ 
is the Eddington mass accretion rate. $\dot{M}_\mathrm{Edd}$ is set as 
the mass accretion rate at which the BH radiates at the Eddington 
luminosity if the radiative efficiency is $\epsilon=10\%$. The Eddington 
luminosity is given by
\begin{equation}
    L_\mathrm{Edd}=\frac{4\pi G M_{\rm BH}c}{\kappa}\sim 1.26\times10^{39}\left(\frac{M_{\rm BH}}{10\text{M}_{\odot}}\right)\text{erg}\,\text{s}^{-1}
\end{equation}
with $G$ the gravitational constant, $M_{\rm BH}$ is the mass of the BH, 
and $\kappa$ is the opacity. A typical value for the case of Thomson 
scattering of free electrons is $0.4\,\text{cm}^2\,\text{g}^{-1}$. This 
gives
\begin{equation}
    \dot{M}_\mathrm{Edd}=\frac{L_\mathrm{Edd}}{\epsilon c^2}\sim 2.19\times 10^{-7}\left(\frac{M_{\rm BH}}{10\text{M}_{\odot}}\right)\text{M}_{\odot}\,\text{yr}^{-1}.
\end{equation}
In massive systems where the stellar companion to the BH fills its Roche 
lobe, the high mass transfer rate may lead to a super-Eddington accretion 
($\dot{M}_{\rm net}/\dot{M}_\mathrm{Edd}$ > 1) and the formation of 
ultraluminous X-ray binaries \citep[e.g.,][]{quast2019}. When the stellar 
companion to the BH does not fill its Roche lobe, the BH can accrete mass 
from the stellar wind of the companion. In this case, an accretion disk 
can form if the specific angular momentum of the accreted stellar wind 
matter is sufficient to orbit around the BH beyond its event horizon.

\subsection{Sub-Eddington accretion regime}

When an accretion disk forms around a BH with a stellar companion that 
does not fill its Roche lobe, the mass accretion rate typically lies 
between 0.1\% and 100\% of the Eddington mass accretion rate 
\citep[e.g., see Fig.\,A.1 of][]{Sen2021}. Accretion is mediated by a 
geometrically thin and optically thick accretion disk. The radiative 
efficiency lies $\epsilon$ between $6\,\%$ and $43\,\%$, depending 
on the black hole spin and the misalignment between the disk and the 
black hole spin \citep{Novikov1973}. 

A decrease in the mass accretion rate without a change in the accretion 
flow geometry is possible, for instance, if the density of the OB star 
wind drops without changes in the wind velocity. In this case, the plasma 
might no longer be dense enough to cool radiatively at a rate which 
balances viscous heating. The disk thickens and becomes a `slim-disk', 
as the relative contribution of the thermal support increases. A large 
fraction of the gravitational potential energy released might be advected 
into the black hole's event horizon, as described by the ADAF 
\citep{Ichimaru1977,Narayan1994}. Therefore, the radiative efficiency 
$\epsilon$ drops. It might be the reason for the transition between the 
high/soft and the low/hard states we observe in black hole hosting X-ray 
binaries \citep{Belloni2010}. It could also explain the low X-ray 
luminosity of the BH candidate around the O-type main-sequence star 
HD96670, in spite of its short orbital period \citep{Gomez2021}.

Alternatively, if the flow is magnetized, it can also transit to a 
magnetically-arrested disk (MAD, \citealp{Begelman2022}) where 
accretion proceeds through episodic ejection of magnetic bubbles 
via the interchange instability \citep{Porth2021} and through a 
reconnecting current sheet between the inner edge of the disk 
and the black hole event horizon \citep{Ripperda2022}. On the 
other hand, if the mass supply rate decreases due to an increase 
in wind speed and/or orbital separation, it is accompanied by a 
drop in specific angular momentum and the disk vanishes. 

Owing to the scale-invariant properties of black holes, emission 
from supermassive black holes can provide fruitful insights to 
model X-ray faint stellar-mass black holes. Sagittarius A$^*$ is 
thought to be fed by the winds from orbiting nearby Wolf-Rayet 
stars \citep{Russell2017,Ressler2023}. Only a fraction of the 
mass supplied at the Bondi radius reaches the black hole event 
horizon though: Faraday rotation \citep{Bower2003,Marrone2007,Wang2013} 
and extrapolation of magneto-hydrodynamic simulations \citep{Ressler2018} 
lead to $\dot{M}_{\rm net}/\dot{M}_\mathrm{Edd}\sim 10^{-8}-10^{-6}$ 
in the immediate vicinity of the BH, but the presence of a disk 
is unclear. In spite of its status of active galactic nucleus, 
M87$^*$ also accretes at a low rate, with 
$\dot{M}_{\rm net}/\dot{M}_\mathrm{Edd}\sim 10^{-6}$ \citep{Prieto2016}, 
orders of magnitude below the geometrically-thin, optically-thick 
disk regime. 

It was found from polarimetry that the emission from these objects 
originates in a dilute and highly magnetized corona populated with 
hot synchrotron-emitting electrons \citep{Bower2018}. Similar 
conclusions were drawn for stellar-mass black holes accreting at 
a rate below $10^{-2}\dot{M}_\mathrm{Edd}$ (e.g. Cygnus X-1 in the 
low/hard state, \citealp{Cangemi2021}). The range of mass accretion 
rates onto stellar mass BHs from the stellar wind of its companion 
is also expected to be the range 
$\dot{M}_{\rm net}/\dot{M}_\mathrm{Edd}\sim 10^{-7}$ - 10$^{-2}$ 
\citep[Fig. A.1 of][]{Sen2021}. In both stellar-mass and supermassive 
BHs, coronal heating is ensured by particle acceleration processes 
at play once the plasma becomes dilute enough to be collisionless 
\citep{Gruzinov1998}. 

\subsection{Particle acceleration}
\label{sec:acceleration}
In collisionless environments, particles are susceptible to being 
accelerated up to relativistic speeds by shocks, turbulence and 
magnetic reconnection \citep{Gruzinov1998}.

\subsubsection{Shocks and turbulence}
The surroundings of an accreting black hole are prone to magnetized 
relativistic shocks due to the accretion-ejection dynamics. They are 
expected in black hole's jets \citep{Malzac2013} and collimated disk 
outflows \citep{Jacquemin-Ide2021}. At shocks, particles can bounce 
back and forth across the shock and gain each time a certain amount 
of kinetic energy \citep[see review by][]{Pelletier2017}. This process 
called diffusive shock acceleration (i.e. first-order Fermi process), 
can accelerate particles up to relativistic speeds \citep{Ellison1990}. 
Furthermore, black holes' corona may show strong Alfvénic turbulence 
\citep{Sandoval2023,Groselj2024} where particles undergo stochastic 
acceleration (i.e., second-order Fermi process) through multiple 
magnetic mirroring episodes which provide them with a net amount of 
kinetic energy \citep{Fermi1949}. Both processes accelerate ions and 
electrons into a non-thermal power-law energy distribution 
\citep{Comisso2022}. It has been shown to be a process susceptible 
to contributing to the heating of black holes' coronae 
\citep{Chandran2018,Hankla2022}.

\subsubsection{Magnetic reconnection}
Particles can also be accelerated via magnetic reconnection which 
occurs in current sheets formed at the interface between magnetic 
field lines of opposite polarity. In these sheets, a very high 
non-ideal electric field appears at the X-points when magnetic 
field lines reconnect \citep{Asenjo2019}. It accelerates particles 
which gather into magnetic islands, called plasmoids, formed in the 
current sheet by the tearing instability \citep{Zenitani2007}. The 
rate at which electromagnetic energy is dissipated and converted 
into particle kinetic energy (the reconnection rate) depends on the 
cold magnetization parameter $\sigma$ of the plasma, defined as the 
ratio of the magnetic to the inertial mass energy of the particles 
\citep[for a recent review, see][]{Kagan2015}:
\begin{equation}
\label{eq:magn}
    \sigma=\frac{B^2}{4\pi n\Gamma_{\rm L} m_e c^2}
\end{equation}
with $B$ the magnitude of the magnetic field, $n$ the plasma number 
density, $\Gamma_{\rm L}$ the bulk Lorentz factor and $m_e$ the mass 
of the electron. Estimates for stellar-mass black holes indicate 
that the magnetic field in which the electrons are embedded can be 
strong enough to accelerate them up to relativistic Lorentz factors 
via magnetic reconnection in the corona \citep{Poutanen2009}. The 
efficiency of this mechanism is hardly affected by the presence or 
not of a disk, as shown by the spherically-symmetric simulations 
ran by \citep{Galishnikova2023} where reconnecting current sheets 
spontaneously form as the highly magnetized plasma is accreted. 
Furthermore, the reconnection rate is enhanced by turbulence 
\citep{Lazarian1999,Lazarian2012} and turbulence-driven magnetic 
reconnection has been shown to be an efficient acceleration 
mechanism at the basis of the black hole's jet \citep{Singh2015}.

\subsubsection{Relative contribution}
\label{section_relative_contribution}

The relative contribution of the above mechanisms to particle 
acceleration depends on the magnetization parameter $\sigma$. 
For instance, particle acceleration at relativistic shocks is 
significantly quenched when $\sigma$ is above $10^{-2}$ 
\citep{Lemoine2010,Sironi2013,Plotnikov2018}, while the magnetic 
reconnection rate plateaus at 10\% when $\sigma$ is above a few 
10 \citep{Werner2018}.

We now evaluate the magnetization parameter in the BH vicinity 
assuming that the stellar magnetic flux is advected inward by 
the accreted material. We work in the wind accretion regime, 
where the wind speed $\upsilon_{\rm w}\gg$ $a\Omega$ the orbital 
speed ($\Omega$ is the orbital angular speed) and the accretion 
radius $R_\mathrm{acc}\ll a$ the orbital separation. At the bow 
shock, formed upstream of the black hole by the gravitational 
beaming of the stellar wind, the density can be approximated by:
\begin{equation}
    \rho = \frac{\dot{M}_{\rm w}}{4\pi a^2 \upsilon_{\rm w}}
\end{equation}
where $\dot{M}_{\rm w}$ is the stellar mass loss rate. Similarly, 
if the stellar magnetic field is dominated by its radial component 
\citep{ud-Doula2002}, the magnetic field at the bow shock is:
\begin{equation}
    B = B_* \left(\frac{a}{R_*}\right)^{-2}
\end{equation}
where $R_*$ is the stellar radius and $B_*$ is the magnetic 
field at the stellar surface. Within the shock, we assume 
spherical Bondi accretion and get the mass density and magnetic 
field profiles \citep{Cunningham2012}:
\begin{align}
    &\rho\propto r^{-(1.5-s)}\\
    &B\propto r^{-2}
\end{align}
where we introduced the parameter $s$ to represent the decrease 
of the accretion rate as we get closer to the BH due to outflows 
\citep{Xie2012}. Hereafter, in Sect.\,\ref{sec:disk_formation}, 
we introduce the accretion radius which stands for the gravitational 
cross-section of the BH. As the stellar wind is beamed toward 
the BH, it forms a bow shock of size comparable to the accretion 
radius. Within this region, we assume that the plasma flowing 
onto the BH is spherically symmetric. Then, using equation\,\eqref{eq:magn} 
and the expression the accretion radius in equation\,\eqref{eqn_Racc}, 
we deduce the magnetization parameter $\sigma$ of the electrons 
at $10r_g$, where $r_g=G M_{\rm BH}/c^2$ is the gravitational 
radius from the BH, to be:
\begin{equation}
    \sigma=5^{-2.5-s}\frac{m_p}{m_e}\frac{B_*^2a^2\upsilon_{\rm w}}{\dot{M}_{\rm w}c^2}\left(\frac{R_*}{a}\right)^{4}\left(\frac{c}{\upsilon_{\rm w}}\right)^{5+2s}
\end{equation}
where we assumed that $n=\rho/m_p$ since the ions' mass dominates 
and $\Gamma_{\rm L}$ = 1. We consider a fiducial value of $s=0.4$ 
to account for the outflows within the shocked region around the 
BH \citep{Xie2012}. In these conditions, lower limits on $\sigma$ 
can be obtained by setting $\upsilon_{\rm w}$ to the terminal wind 
speed of an O star (typically $\sim$2,000\,km$\,$s$^{-1}$). We take 
the following standard values: a stellar mass loss rate typical of 
an LMC O-type star $\dot{M}_{\rm w}=10^{-7}\,$M$_{\odot}\,$yr$^{-1}$ 
\citep{Brands2022}, a stellar magnetic field of $B_*=30$\,G 
\citep[30 G is below the detection sensitivity of $\sim$80\% of 
the stars observed in the MiMeS survey,][]{Wade2016,Grunhut2017,Petit2019}, 
and a stellar radius of $R_*=20\,$R$_{\odot}$. Then, for orbital 
separations ranging between 3\,R$_*$ and 100\,R$_*$, $\sigma$ is 
at least $10^3-10^6$. The magnetization parameter can also be 
estimated from observations interpreted via radiative models. For 
instance, in the BH-hosting HMXB Cygnus X-1, the analysis of the 
polarized synchrotron emission \citep{Cangemi2021} indicates that 
$\sigma=10^3-10^6$ in the innermost regions, which is coherent 
with the values obtained for a magnetic field near equipartition 
with the ion energy density \citep{Malzac2009}. This agrees with 
the values we derived from our toy model of stellar magnetic flux 
compression.

Given the high magnetization parameters we can expect around stellar-mass 
BHs fed by the wind from an OB stellar companion, it is safe to assume 
that the main particle acceleration mechanism in the corona is magnetic 
reconnection in the relativistic regime (i.e. $\sigma$ higher than a few 
10), where the reconnection rate is $\sim$10\% (or even higher if 
reconnection is turbulence-driven). In the MAD regime, the approximate 
equipartition between accretion and magnetic energy density implies that 
a large fraction of $\dot{M}_{\rm net}c^2$ could serve to accelerate the 
electrons via magnetic reconnection. This is in agreement with observations 
of supermassive BHs where fits from detailed radiative models indicate that 
the electron heating parameter $\delta$, that is, the fraction of the 
viscously dissipated energy that heats the electrons, is high, with 
values ranging between $0.1$ and $0.5$ \citep{Yuan2003,Yu2011,Liu2013}.

\subsection{Non-thermalised emission}
\label{sec:}

In classic radiatively-inefficient accretion flows, be it because of 
the lack of accretion disk or of the low $\dot{M}_\mathrm{net}/\dot{M}_\mathrm{Edd}$, 
the ions and the electrons are no longer coupled through Coulomb 
interactions due to the low plasma density \citep{Moscibrodzka2016}. 
The temperatures of the ions and electrons differ and the electron 
distribution can develop a significant non-thermal component through 
particle acceleration. Indeed, Coulomb collisions between electrons 
become negligible for $\dot{M}_\mathrm{net}/\dot{M}_\mathrm{Edd}\ll 0.06$ 
\citep{Sharma2007}, hence the presence of a non-thermal population to 
account for the spectrum of Cygnus X-1 in the low/hard state for instance 
\citep{Cangemi2021}. We have shown in Sect.\,\ref{sec:acceleration} 
that the energy reservoir contained in the magnetized accretion flow 
can be efficiently tapped through magnetic reconnection in the very 
sub-Eddington regime $\dot{M}_\mathrm{net}/\dot{M}_\mathrm{Edd}\ll 1$ 
(see also \citep{Sharma2007}). The electrons accelerated through this 
process will emit non-thermal radiation through synchrotron, bremsstrahlung, 
and inverse Compton scattering. Hereafter, we focus on the collisionless 
regime and estimate the synchrotron luminosity from the purely non-thermal 
population of electrons which should dominate at low mass accretion rate.

\subsubsection{Synchrotron}
\label{section_synchrotron}

As charged particles spiral around magnetic field lines, they emit 
synchrotron radiation. The radiative energy emission rate from a 
single electron of charge $-e$ (with $e>0$) and Lorentz factor 
$\gamma$ in an ambient magnetic field $B$ is:
\begin{equation}
    L_\mathrm{sync,1}=\frac{4}{3}\gamma^2 c \sigma_T\frac{B^2}{8\pi}
\end{equation}
where $\sigma_T$ is the electron cross-section for Thomson scattering. 
The total synchrotron luminosity $L_\mathrm{sync}$ produced by the 
electrons contained in a fiducial uniform sphere of radius $10r_g$ 
around the BH is:
\begin{equation}
    L_\mathrm{sync}=\frac{2}{9}(10r_g)^3c\sigma_T n\gamma^2 B^2
\end{equation}
where $n$ is the number density of electrons and $r_g=GM_{\rm BH}/c^2$. 
We obtain a synchrotron luminosity:
\begin{equation}
    L_\mathrm{sync}\sim 10^{35}\left(\frac{n}{\text{$10^{11}$cm$^{-3}$}}\right)\left(\frac{B}{10^6\text{G}}\right)^{2}\left(\frac{\gamma}{100}\right)^{2}\left(\frac{M_{\rm BH}}{20\text{M$_{\odot}$}}\right)^{3}\text{erg}\,\text{s}^{-1}
\end{equation}
where we used the ambient magnetic field and electron number density 
values in the denominator found by \citet{Cangemi2021} in the vicinity 
of the $\sim$20M$_{\odot}$ BH in Cygnus X-1 \citep{Miller-Jones2021}, 
from the analysis of the polarized emission.

This idealized one-zone model does not account for the non-uniform 
density and magnetic field, nor the shape of the underlying distribution 
of electron energy. In the relativistic regime, the maximum Lorentz 
factor reachable by an electron accelerated by magnetic reconnection 
over a length scale $r_g$ in an ambient magnetic field $B$ is:
\begin{equation}
    \gamma_\mathrm{max}\sim\frac{eBr_g}{m_ec^2}\sim K \sigma
\end{equation}
where $K$ is the plasma multiplicity. Even at low mass accretion 
rates and for a maximally rotating BH ($\bar{s}=1$), we expect 
$K>1$ \citep{ElMellah2022}, so $\gamma_\mathrm{max}>10^3$ in the 
magnetized plasma surrounding stellar-mass BHs capturing the wind 
from an OB companion. 

Acceleration through magnetic reconnection is typically described 
by a power law distribution $N(\gamma)\propto \gamma^{-p}$ with an 
exponent $p\sim 0.9-1.2$ \citep{Werner2016,ElMellah2022}. Hence, 
the contribution of the highest energy electrons dominate the 
emission. However, this simplified estimate of the synchrotron 
power with $\gamma=100$ shows that, provided electrons are 
accelerated to significant Lorentz factors and embedded in a 
high magnetic field, they can produce a significant amount of 
non-thermal synchrotron emission. For stellar-mass BHs without 
an accretion disk, the peak of this emission is expected at a 
photon energy of 17\,keV (with the values above of Lorentz 
factor and magnetic field), in hard X-rays.

\subsubsection{Bremsstrahlung}
In the absence of magnetic fields, particles free fall directly in 
the BH such that energy transfer and cooling are negligible. In this 
case, the X-ray emission is bremsstrahlung dominated and the process 
can be treated adiabatically. From the mass continuity equation 
$\dot{M}_{\rm acc} = 4\pi r^2 \rho(r) \upsilon(r)$, we can write
\begin{equation}
    \rho(r) = \dot{M}_{\rm acc} / 4\pi r^2 \upsilon(r),
\end{equation}
where $r$ is the distance to the BH. Assuming $\upsilon(r) = \sqrt{2GM_{\rm BH}/r}$ 
as the free-fall velocity, we get 
\begin{equation}
    \rho(r) = \frac{\dot{M}_{\rm acc}}{\sqrt{32\pi^2GM_{\rm BH}r^3}}.
\end{equation}

For an adiabatic process, $T \propto \rho^{2/3}$, the temperature 
profile of the free-falling particles become
\begin{equation}
    T(r) = T_{0}\left(\frac{\dot{M}_{\rm acc}}{\sqrt{32\pi^2GM_{\rm BH}r^3} \rho_{0}}\right)^{2/3}
\end{equation}
where $T_{0}$ and $\rho_{0}$ are assumed to match the ambient wind 
temperature and density of the OB star companion, far away from the 
BH. The energy per unit time emitted by the gas into a $4\pi$ solid 
angle from a volume $dV$ is given by \citep{Frank2002}
\begin{equation}
    L_{\rm brem}(r) \propto \int_{R_{\rm sch}}^{R_{\rm acc}} n_{\rm e}(r) \cdot n_{\rm p}(r) \cdot T^{1/2}(r) \cdot dV
    \label{lum_brem}
\end{equation}
where $n_{\rm e}(r)$ and $n_{\rm p}(r)$ are electron and proton number 
densities respectively and $T(r)$ is the temperature stratification. 
Assuming the infalling wind matter is fully ionised, spherically 
symmetric, and composed of hydrogen and helium, Eq.\,(\ref{lum_brem}) 
can be integrated from the Schwarzschild radius $R_{\rm sch}$ to the 
accretion radius $R_{\rm acc}$ using the temperature and density 
dependencies above. 

For a stellar-mass BH accreting mass from the interstellar medium, it 
has been shown that the resulting luminosity is extremely low, orders 
of magnitude below observable limits (see discussion in section.\,7.8, 
following equation.\,(7.16) of \citealp{Frank2002}). For BHs accreting 
material from the wind of an OB star companion, the mass accretion rate 
is largely set by the wind mass loss rate of the OB star and the binary 
orbital period. The radiative luminosity in the X-ray band is at most 
$\sim10^{30}$\,erg$\,$s$^{-1}$ at orbital period of 10\,d and falls 
steeply with the mass accretion rate, such that at orbital periods of 
100\,d the luminosity is less than $\sim10^{27}$\,erg$\,$s$^{-1}$ 
(Quast et al. in prep). Hence, thermal bremsstrahlung is not expected 
to contribute to observable X-ray emission. 

X-ray line emissivity due to bound-bound transitions can be significantly 
higher than the thermal bremsstrahlung emission if the temperature of 
ions drops below 10$^6$K. However, this temperature regime should not 
be relevant near the BH since the ion temperature is much higher than 
the electron temperature in the magnetized plasma, and electron 
temperature is typically above 10$^6$K.

\subsubsection{Inverse Compton scattering}

Relativistic electrons upscatter soft X-ray photons through Inverse 
Compton in optically thin and hot accretion flow. The seed photons 
can come from the multi-colour black body emission from an underlying 
thin disk (at high mass and angular momentum accretion rates), or the 
synchrotron emission. In the latter case, it could shift the peak of 
synchrotron emission up to hard X-rays \citep{Sridhar2021}, although 
Comptonisation is expected to be sub-dominant when 
$\dot{M}/\dot{M}_\mathrm{Edd}\ll 1$ \citep{Esin1997}.


\section{Method}
\label{section_method}

We discuss the grid of binary evolution models utilised for this work 
in Sect.\,\ref{section_models}. We discuss the criterion for forming 
an accretion disk around the BH in BH+OB star binaries in 
Sect.\,\ref{sec:disk_formation}. We outline the procedure to calculate 
the X-ray luminosity from a BH+OB star binary in the presence and 
absence of a geometrically-thin, optically-thick accretion disk in 
Sect.\,\ref{section_acc_disk_present} and Sect.\,\ref{section_acc_disk_absent}, 
respectively. In Sect.\,\ref{section_distribution_functions}, we define 
the distribution functions that are used to predict the observable stellar 
parameters from our binary models during the BH+OB star phase. 

\subsection{Stellar evolution models}
\label{section_models}

We use the detailed binary evolution models computed by \citet{pablothesis} 
and \citet{Pauli2022} using the 1D stellar evolution code 
MESA\footnote{\href{https://docs.mesastar.org/en/release-r23.05.1/index.html}{https://docs.mesastar.org/en/release-r23.05.1/index.html}} 
\citep[Modules for Experiments in Stellar Astrophysics,][]{mesa11,mesa13,mesa15,mesa18,mesa19}, 
version 8845 and 10398 respectively\footnote{inlists can be found at 
\href{https://github.com/orlox/mesa_input_data/tree/master/2016_binary_models}{this website}}. 
A detailed description of all stellar and binary physics assumptions can 
be found in \citet{pablothesis}, \citet{Langer2020}, \citet{Sen2022}, 
\citet{Pauli2022} and \citet{Sen2023}. We outline below the 
necessary details required to follow our work. 

The initial mass of the primary star $M_{\rm 1,i}$ (initially more massive 
and forms the compact object) ranges from $\sim$10-90\,$M_{\odot}$ in steps 
of log\,($M_{\rm 1,i}$/$M_{\odot}$) = 0.05. The initial 
orbital periods $P_{\rm i}$ and mass ratios $q_{\rm i}$ (initial mass of the 
secondary star $M_{\rm 2,i}$ divided by the initial mass of the primary star 
$M_{\rm 1,i}$) range from 1.4-3162\,d (in steps of log\,($P_{\rm i}/d$ = 0.05) 
and 0.25-0.95 (in steps of 0.05), respectively. Models with initial primary 
masses 10-40\,$M_{\odot}$ are taken from the work of \citet{pablothesis}, 
and the models with initial primary masses 40-90\,$M_{\odot}$ are taken 
from the work of \citet{Pauli2022}. Both sets of models assume a metallicity 
suitable for the LMC. 

The models start from the onset of core hydrogen burning of the primary, 
and both components are assumed to start core hydrogen burning simultaneously. 
When the initially more massive star fills its Roche lobe, mass transfer 
via Roche-lobe Overflow is modelled using the ``contact" scheme in MESA 
\citep{Marchant2016}. Mass transfer is assumed to be conservative until 
the accretor spins up to critical rotation \citep{packet1981}. Any further 
mass transferred from a Roche-lobe filling donor to a critically rotating 
stellar companion is removed from the star through enhanced stellar wind 
mass and momentum loss \citep{langer2003,petrovic2005}. When the combined 
luminosity of both binary components is insufficient to drive this excess 
mass loss, we assume the binary model will merge and terminate the evolution. 

If both binary components fill its Roche lobe during a mass transfer phase, 
the evolution of the binary model in such a contact configuration is 
calculated until one of the stars overflows the L2 Lagrangian point of 
the binary model. Otherwise, the evolution of both stars is followed until 
the end of the core carbon burning of the primary. The details of the mass 
transfer model and the ensuing mass transfer efficiency during different 
mass transfer phases are studied in detail in \citet{Sen2022}. In the models, 
mass transfer is conservative until orbital periods of 5\,d, where tides 
\citep{detmers2008,zahn1977} can halt the spin-up of the mass-gaining star. 
For longer orbital periods, the overall mass transfer efficiency of the 
binary models is $\sim$5-10\%. 

Upon core carbon depletion of the primary, we assume that if the helium core 
mass of the primary is larger than 6.6\,$M_{\odot}$, the helium core of the 
primary directly collapses into a BH without a natal kick and the mass of the 
BH formed equals the helium core mass of the primary \citep{Ertl2016,Muller2016,Sukhbold2018,Langer2020}. The BH kick 
depends on the neutrino energy available and the asymmetry of the fall-back 
material \citep{Belczynski2012,Janka2013}. Empirical evidence towards the 
magnitude of BH kicks remains inconclusive \citep[e.g.,][]{Farr2011}, with 
some studies positing the need for a low kick \citep{Wong2012}, while others 
requiring a significantly high kick \citep{Repetto2012}. It has been recently 
shown that a BH kick is not necessary to explain the lack of observed wind-fed 
BH HMXBs in the Milky Way compared to the number of Wolf-Rayet+O star binaries 
\citep{Sen2021}. When the first BH forms, the secondary (hereafter called the 
OB star companion to the BH) still burns hydrogen in its core. This marks the 
onset of the BH+OB star phase studied in this work. 

The further evolution of the OB star companion in the BH+OB binary is 
modelled as a single star. The mass and angular momentum loss determine 
the orbital period evolution during the BH+OB star phase via the stellar 
wind of the OB star companion \citep{quast2019,ElMellah2020}. For the 
typical mass ratios of the BH to the OB star companion, and the fraction 
of the OB stellar wind accreted by the BH \citep{Sen2021}, we can assume 
that the orbital period remains constant during the BH+OB star phase 
\citep[][Fig.\,10]{ElMellah2020}. Hence, we assume the orbital period 
during the BH+OB star phase is equal to the orbital period of the binary 
model at the formation of the BH. The above simple assumptions result 
in a strict lower limit to the BH mass, and a constant orbital period 
during the BH+OB star phase. This leads to a small modelling uncertainty 
(at most a factor of 2, \citealp{ElMellah2020}) in the calculated mass 
accretion rate $\dot{M}_{\rm acc}$ (Eq.\,(\ref{equation_mass_accretion_rate})). 

Since the models undergo mass transfer before the BH+OB star phase and 
there is no kick velocity, we assume the orbit during the BH+OB star 
phase will remain circular. Lastly, we assume that the BHs formed from 
the collapse of the primary have negligible spins (Kerr parameter $\chi$ 
= 0, \citealp{Qin2018,Ma2023}). The BH+OB star phase ends when the OB 
star companion completes core hydrogen burning or fills its Roche lobe 
while on the main sequence. 

\subsection{Disk formation criterion}
\label{sec:disk_formation}
To determine whether a disk forms, the specific angular momentum of the 
infalling matter onto the BH must be compared to the specific angular 
momentum $l_\mathrm{ISCO}$ of a test particle at the ISCO of the BH. 
For a spinning black hole, it is entirely set by the black hole mass 
$M_{\rm BH}$ and its dimensionless spin $\bar{s}\in [0,1]$, with 
$\bar{s}=0$ for a non-spinning black hole and $\bar{s}=1$ for a 
maximally spinning black hole: 
\begin{equation}
    l_\mathrm{ISCO}=\pm\frac{GM_{\rm BH}}{c}f(\bar{s})
\end{equation}
where 
$f(\bar{s})$ a dimensionless function of $\bar{s}$ only. Efficient angular 
momentum transport predicts that the first-born BH in a massive binary 
system will be slowly spinning \citep{Qin2018,Ma2023,Marchant2023}. 

If accretion proceeds through stellar wind capture (instead of Roche lobe 
overflow), a bow shock forms around the black hole \citep{ElMellah2015}. 
Within the shocked region, matter flattens into a disk, that is, a 
centrifugally-maintained structure, provided its specific angular momentum 
is higher than $l_\mathrm{ISCO}$. When the BH accretes matter from the OB 
star wind, and the wind speed at the orbital separation $\upsilon_{\rm w}$ 
is larger than the orbital speed $a\Omega$, we have (c.f. equation (10) of 
\citealp{Sen2021}, where the criterion is expressed in terms of the ratio 
of the circularisation radius of a Keplerian accretion disk to the innermost 
stable circular orbit of the BH)
\begin{equation}
\label{eq:circ}
    \frac{l}{l_\mathrm{ISCO}}\sim \frac{2\eta}{(1+q)f(\bar{s})}\frac{c}{a\Omega}\left(\frac{a\Omega}{\upsilon_{\rm w}}\right)^4
\end{equation}
where $a$ is the orbital separation, $\Omega$ is the orbital angular speed, 
$q=M_{\rm OB}/M_{\rm BH}$ is the mass ratio, $M_{\rm OB}$ is the mass of 
the OB star and $\eta$ is the specific angular momentum of the accreted 
material in units of $R_\mathrm{acc}^2\Omega/2$ \citep{shapiro1976}, with 
$R_\mathrm{acc}$ the accretion radius which is given by, in the wind 
accretion regime
\begin{equation}
    R_\mathrm{acc}=\frac{2GM_{\rm BH}}{\upsilon_{\rm w}^2}.
    \label{eqn_Racc}
\end{equation}

Equation\,\eqref{eq:circ} shows that accretion disks form preferentially 
in systems where the wind speed is low compared to the orbital speed. We 
can express the condition for disk formation as 
\begin{equation}
    P_{\rm orb}<4\pi GM_{\rm BH}c \left(\frac{\eta}{f(\bar{s})}\right)\frac{1}{\upsilon_{\rm w}^4}
    \label{eq:disk_formation}
\end{equation}
where $P$ is the orbital period. Hydrodynamical simulations of accretion 
onto stellar-mass BHs suggest that $\eta\sim1/3$ \citep{Livio1986,Ruffert1999}. 
For a Schwarzschild BH, $f(\bar{s})=\sqrt{12}$ \citep{Carroll2019}. In 
the Newtonian approximation, $f(\bar{s})=1$ for a non-spinning BH. In 
Sect.\,\ref{section_results}, we show the results for $f(\bar{s})=1$ to 
be consistent with the accretion disk formation criterion in \citet{Sen2021}, 
but also discuss the statistics for $f(\bar{s})=\sqrt{12}$ in 
Sect.\,\ref{section_accretion_disk_formation}. 

\subsection{X-ray luminosity from a Keplerian accretion disk}
\label{section_acc_disk_present}

If an accretion disk can form, the X-ray luminosity 
$L_{\rm X}$ from the accretion disk is calculated as 
\begin{equation}
    L_{\rm X}=G\frac{M_{\rm BH}\dot{M}_{\rm acc}}{R_{\rm ISCO}},
    \label{eq:Lx}
\end{equation}
where $G$ is the Gravitational constant, $R_{\rm ISCO}$ is the radius 
of the innermost stable circular orbit of the BH, $M_{\rm BH}$ is the 
mass of the BH, $\dot{M}_{\rm acc}$ is the mass accretion rate at the 
accretion radius $R_{\rm acc}$ given by the Bondi-Hoyle accretion rate 
\citep{Bondi1944}, along with a self-limiting effect from the Eddington 
luminosity \citep{Davidson1973,vanbeveren2020} 
\begin{equation}
    \frac{\dot{M}_{\rm acc}}{\dot{M}_{\rm w}} = \eta_{\rm Edd} \frac{\pi R_{\rm acc}^2}{4\pi a^2} \frac{\upsilon_{\rm rel}}{\upsilon_{\rm w}},
    \label{equation_mass_accretion_rate}
\end{equation}
where $\dot{M}_{\rm w}$ is the wind loss rate of the OB star, $a$ is 
the orbital separation, $R_{\rm acc}$ is the accretion radius of the 
BH, $\upsilon_{\rm w}$ is the wind velocity of the O star, 
$\upsilon_{\rm rel}$ is the relative velocity of the BH with respect 
to the wind velocity (see equations (1)-(8) of \citealp{Sen2021}), 
and $\eta_{\rm Edd}$ represents the self-limiting effect of the 
Eddington luminosity,
\begin{equation}
    \eta_{\rm Edd} = \left(1 - \frac{L_{\rm X}}{L_{\rm Edd,BH}}\right)^2.
\end{equation}
Here, $L_{\rm Edd,BH}$ is the Eddington luminosity of the BH given by 
\begin{equation}
    L_{\rm Edd} = \frac{4\pi c G M_{\rm BH}}{\kappa_{\rm e}} = 65335 \frac{M_{\rm BH}}{1+X} \frac{L_{\odot}}{M_{\odot}},
\end{equation}
where $X$ is the hydrogen mass fraction of the accreted material and 
$\kappa_{\rm e} = 0.2(1+X)$ cm$^{2}$ g$^{-1}$ is the electron scattering 
opacity. Combining the above equations, 
\begin{equation}
    L_{\rm X}=\frac{\alpha}{(1+\sqrt{1+\alpha})^2}L_{\rm Edd},
    \label{Lx_eqn}
\end{equation}
where $\alpha$ depends on the Eddington accretion rate $\dot{M}_{\rm Edd,BH}$,
\begin{equation}
    \alpha=4\frac{\bar{\gamma}^2}{(1+\bar{\gamma})^{3/2}}\left(\frac{M_{\rm BH}}{M_{\rm O}}\right)^2 \frac{\dot{M}_{\rm w}}{\dot{M}_{\rm Edd,BH}},
\end{equation}
with 
\begin{equation}
    \bar{\gamma}=\frac{R_{\rm O}}{2a}\left(\frac{\upsilon_{\rm esc}}{\upsilon_{\rm w}}\right)^2, 
\end{equation}
\begin{equation}
    \dot{M}_{\rm Edd,BH} = \frac{L_{\rm Edd} R_{\rm ISCO}}{G M_{\rm BH}},
\end{equation}
\begin{equation}
    R_{\rm ISCO} = \frac{6GM_{\rm BH}}{c^2},
\end{equation}
and 
\begin{equation}
    \upsilon_{\rm esc} = \sqrt{\frac{2GM_{\rm O}}{R_{\rm O}}(1-\Gamma_{\rm e})}
\end{equation}
where $M_{\rm O}$, $R_{\rm O}$ and $\Gamma_{\rm e}$ are the mass, 
radius and Eddington factor of the OB star companion. 

\begin{figure*}
    \includegraphics[width=0.48\hsize]{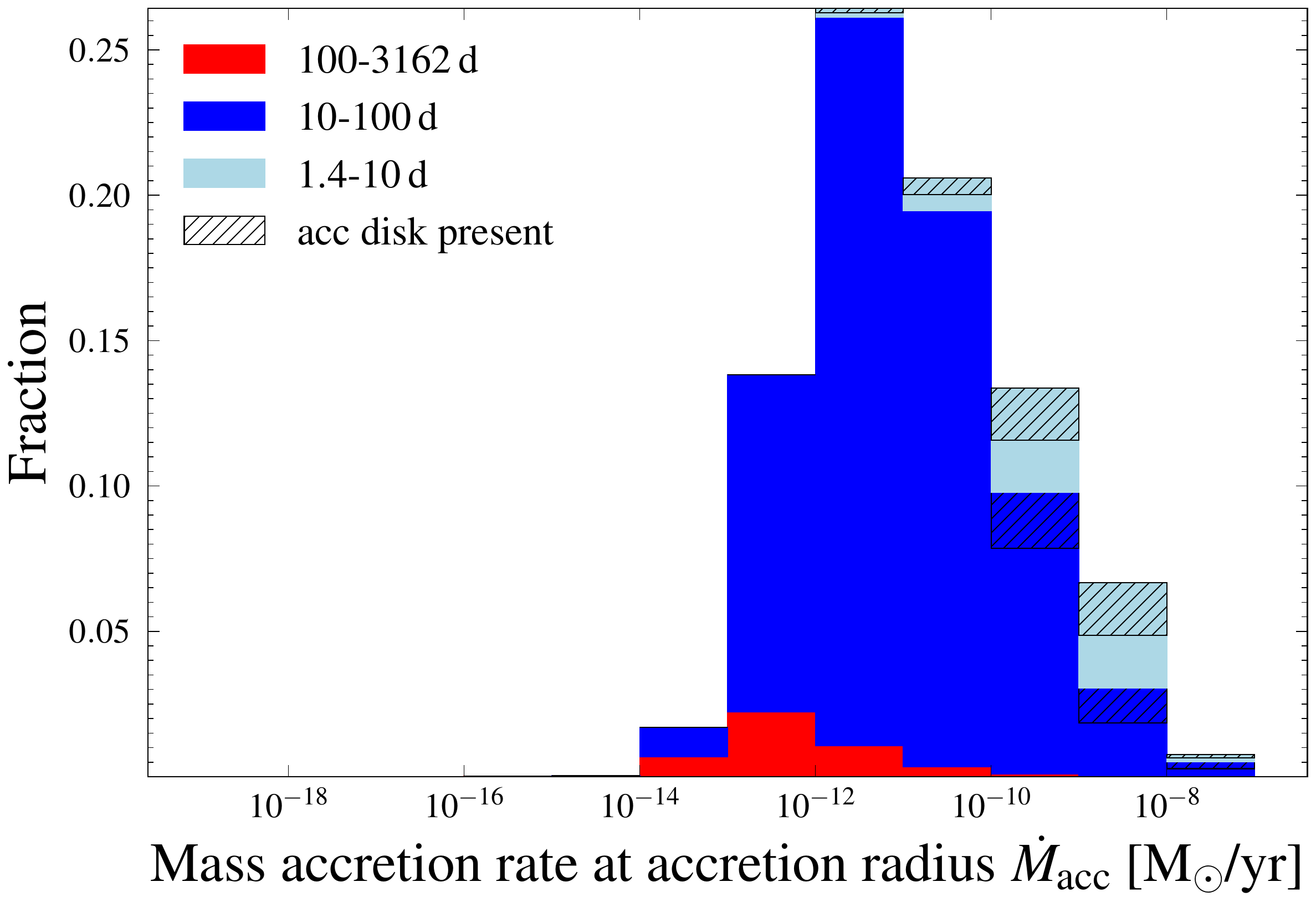}
    \includegraphics[width=0.48\hsize]{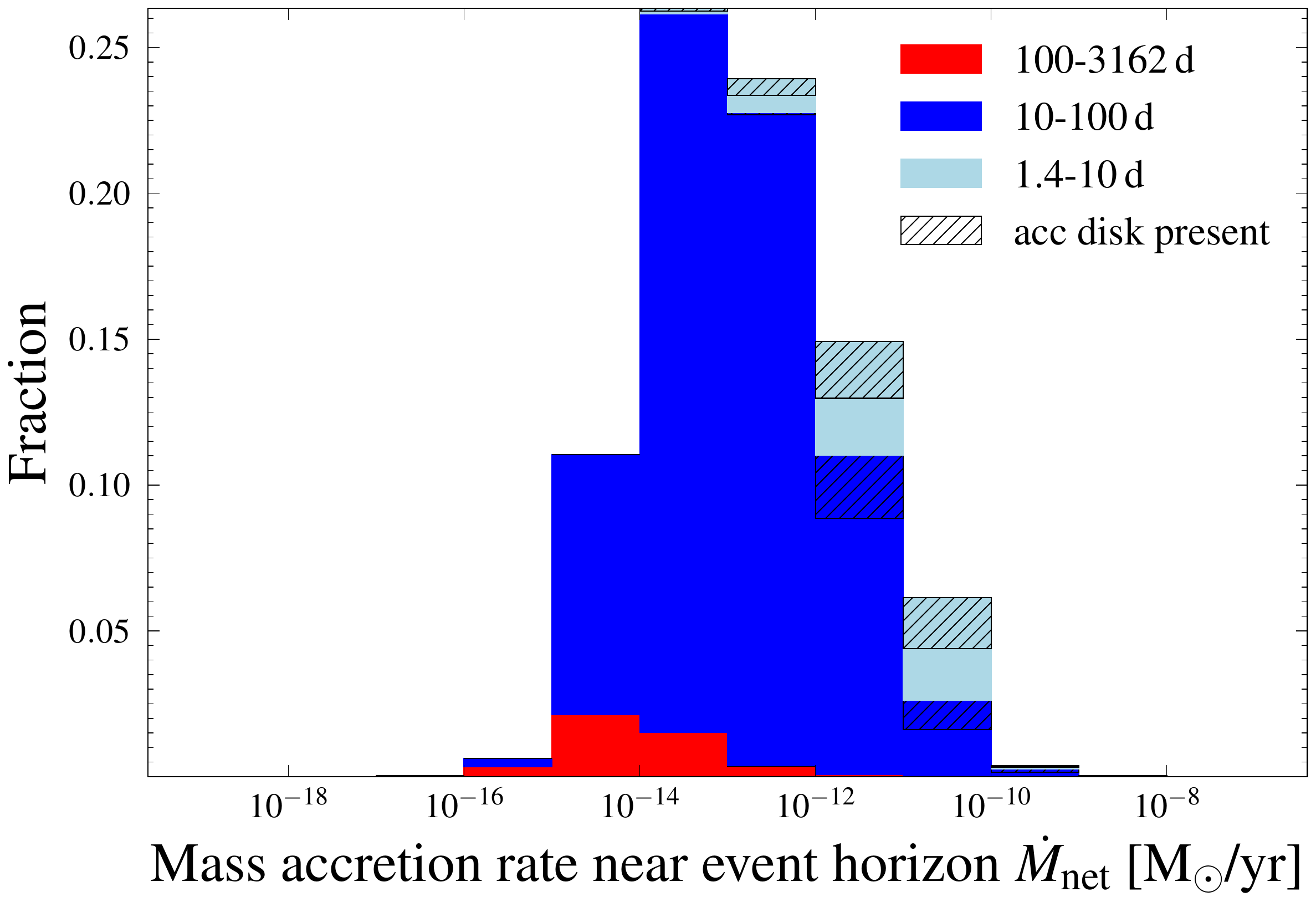}
    \caption{Distribution of the mass accretion rate at the accretion radius (Eq.\,\ref{equation_mass_accretion_rate}, \textit{left panel}) and the mass accretion rate at the event horizon of the BH (Eq.\,\ref{equation_net_mass_accretion_rate}, \textit{right panel}), during the BH+OB star phase. The three colours denote the contributions from different initial orbital period ranges. The coloured histograms are stacked on each other, and the sum of the ordinate values equals unity. The black hatching shows the contribution from the BH+OB star binaries where a Keplerian accretion disk can form. }
    \label{figure_Macc_Mnet}
\end{figure*}

\begin{figure}
    \centering
    \includegraphics[width=0.98\hsize]{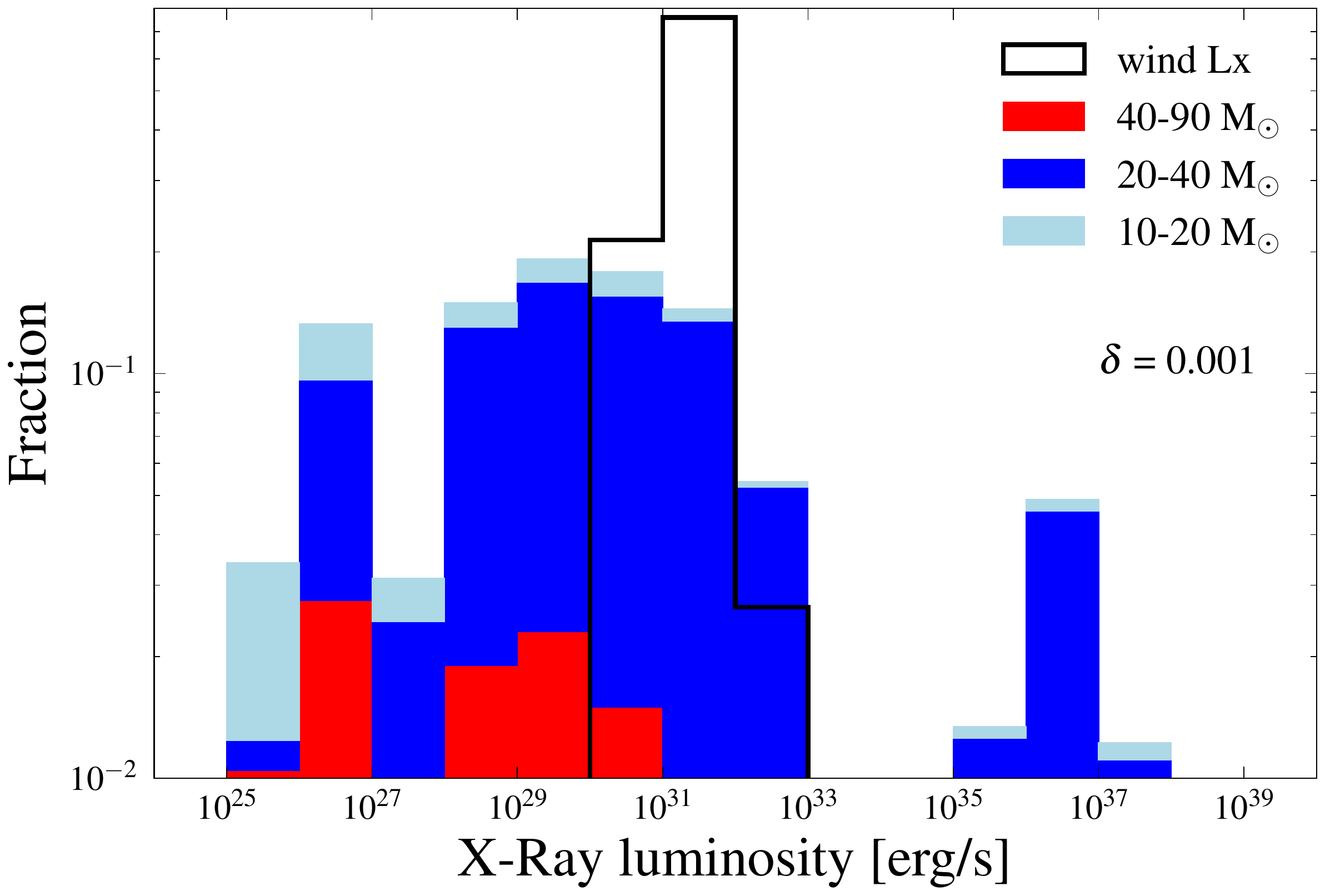}
    \includegraphics[width=0.98\hsize]{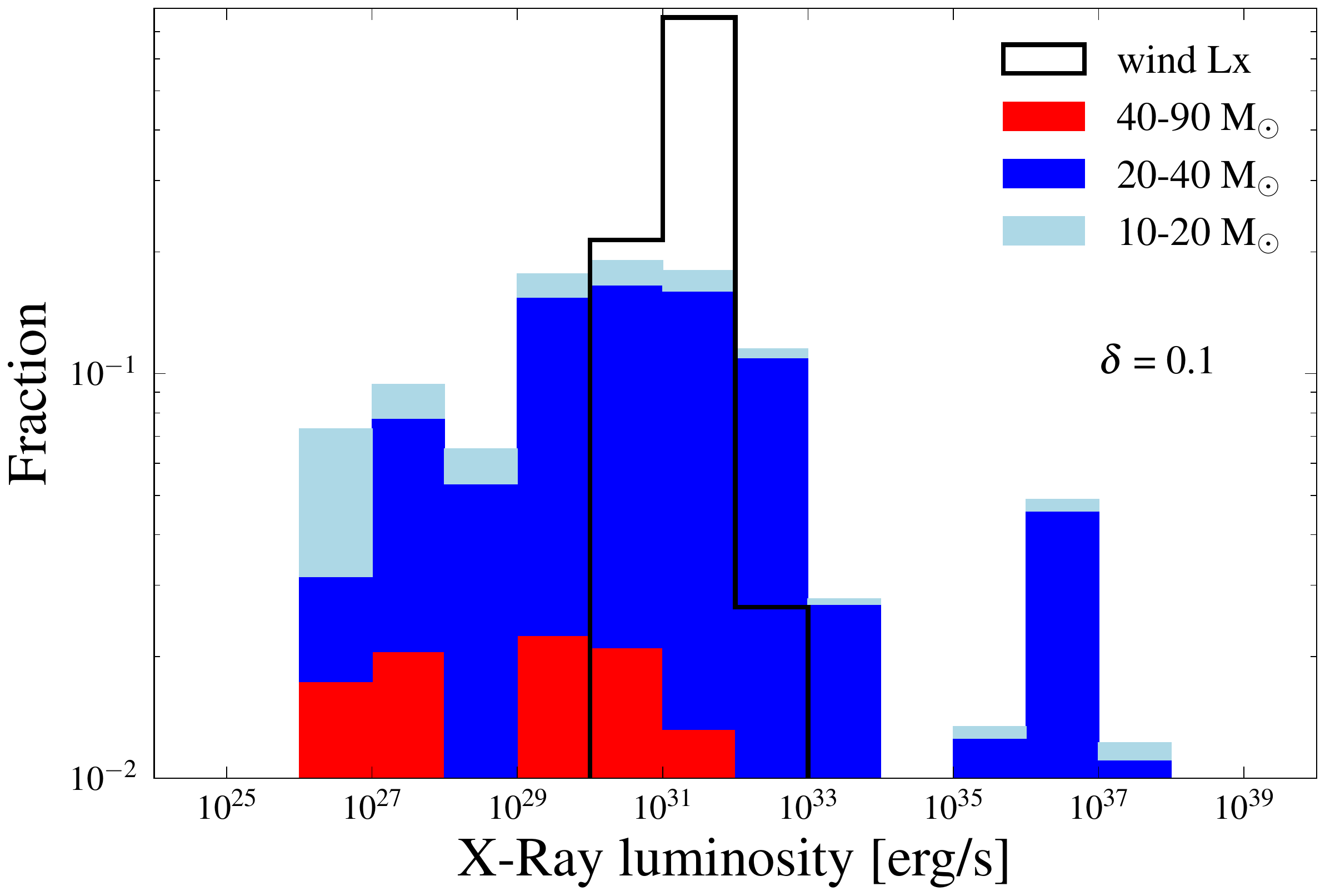}
    \includegraphics[width=0.98\hsize]{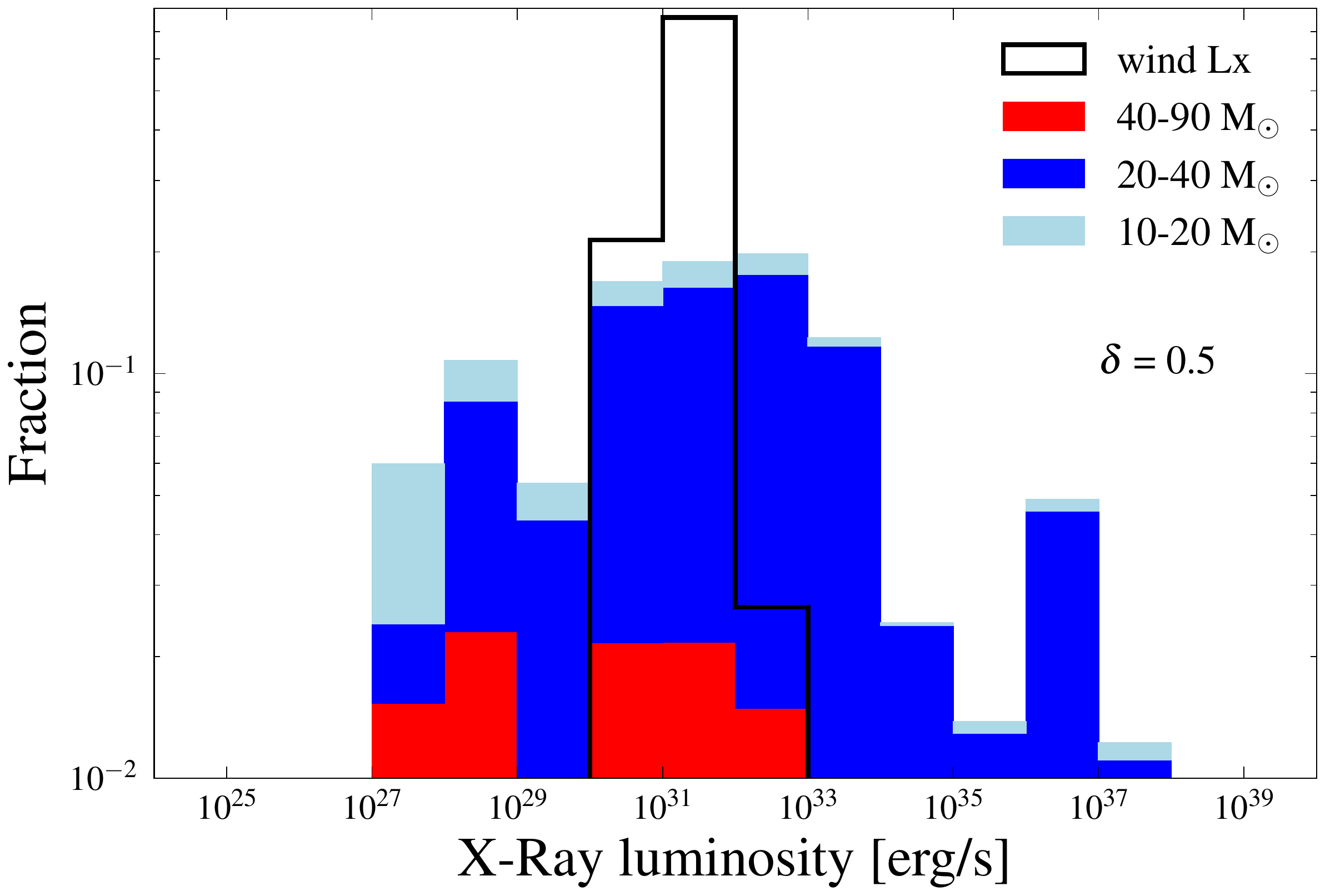}
    \caption{Distribution of X-ray luminosity from BH+OB star binaries for three values of the viscous dissipation parameter $\delta$ = 0.001 (\textit{top panel}), 0.1 (\textit{middle panel}), 0.5 (\textit{bottom panel}). Light blue, dark blue and red colours denote the contributions from binary models having different initial donor mass ranges of 10-20\,$M_{\odot}$, 20-40\,$M_{\odot}$ and 40-90\,$M_{\odot}$ respectively. The coloured histograms are stacked on each other, and the sum of their ordinate values equals unity (see Eq.\,\ref{equation_histogram}). The black step histogram shows the distribution of X-ray luminosity arising from the shocks in the winds of the OB star companion and is separately normalised to unity. }
    \label{figure_Lx}
\end{figure}

\subsection{X-ray luminosity without an accretion disk}
\label{section_acc_disk_absent}

When the specific angular momentum carried by the OB star wind matter 
is insufficient to form a Keplerian accretion disk around the BH, the 
wind matter can spiral into the BH following the magnetic field lines. 
During this in-fall, we have shown in Sect.\,\ref{section_relative_contribution} 
that electrons can receive a portion of the viscously dissipated energy 
of the in-falling matter, and emit radiation in the X-ray band. The 
X-ray luminosity $L_{\rm X}$ is 
\begin{equation}
    L_{\rm X} = \epsilon \dot{M}_{\rm net} c^2,
    \label{equation_Lx}
\end{equation}
where $\epsilon$ is the radiative efficiency and $\dot{M}_{\rm net}$ 
is the net mass accretion rate at the radius of the ISCO of the BH. 

Recent simulations of hot accretion flows have revealed the 
presence of outflows in ADAFs, such that the mass inflow rate of 
the accreting gas decreases as the material approaches the event 
horizon of the BH \citep[][]{Stone1999,Yuan2010}. 
We assume the mass accretion rate scales with the radius from 
the BH by a power-law ($\dot{M} \sim R^s$, 
\citealp{Stone1999,Igumenshchev2000,Stone2001,Yuan2010}). 
Hence, the relation between the net mass accretion rate at 
the radius of the innermost stable circular orbit $\dot{M}_{\rm net}$ 
and the mass accretion rate at the accretion radius $\dot{M}_{\rm acc}$ 
is given by 
\begin{equation}
    \dot{M}_{\rm net} = \dot{M}_{\rm acc} \left( \frac{R_{\rm ISCO}}{R_{\rm acc}} \right)^s.
    \label{equation_net_mass_accretion_rate}
\end{equation}
In consistency with Sect.\,\ref{section_relative_contribution}, we 
assume $s \approx 0.4$ \citep{Xie2012}. We assume this parameter to be 
constant during the BH+OB star phase in all our models. 

\citet[][Fig.\,1]{Xie2012} studied the radiative efficiency $\epsilon$ of hot accretion flows around a stellar-mass BH orbiting an OB star companion. They derive $\epsilon$ as a function of the net mass accretion 
rate $\dot{M}_{\rm net}$ (i.e. $\epsilon$ = $\epsilon$($\delta$,$\dot{M}_{\rm net}$)), for three different values of the viscous dissipation parameter $\delta$ (defined in Sect.\,\ref{section_relative_contribution}). From their figure\,1, 
we extract the values of the radiative efficiency as a function of the 
net mass accretion rate to calculate the X-ray luminosity during the BH+OB 
star phase in our models for viscous dissipation parameter
$\delta = 0.001, 0.1, 0.5$. While we show in Sect.\,\ref{section_radiative_efficiency} 
that viscous heating of electrons can be efficient even in the case of 
stellar mass BHs with OB star companions, we present our results 
for all three values above to derive a lower limit on the number of 
faint X-ray sources from BH+OB star binaries for the most 
inefficient case of viscous heating. 

\subsection{Histogram distribution functions}
\label{section_distribution_functions}

The distribution function of an observable parameter $X_{\rm obs}$ 
is constructed by weighing the values of $X_{\rm obs}$ at each 
timestep during the BH+OB star phase with the initial mass 
function \citep{salpeter1955} and binary distribution functions 
\citep{sana2013} of the progenitor binary model `$\rm m$', then 
summing over all models and normalising to unity. The number fraction 
$h_{\rm obs}$ in a given histogram bin [$X_{\rm 1}$,$X_{\rm 2}$] 
of the observable $X_{\rm obs}$ is given by 
\begin{equation}
    h_{\rm obs} \, ( X_{\rm 1} < X_{\rm obs} < X_{\rm 2} ) = \frac{\sum_{\rm m=1}^{\rm N} \,W_{\rm m} \, \Delta t_{\rm [X_{\rm 1},X_{\rm 2}],m}}{\sum_{\rm m=1}^{\rm N} \, W_{\rm m} \, \Delta t_{\rm BH+O,m}}\:,
    \label{equation_histogram}
\end{equation}
where N is the total number of binary models that go through the 
BH+OB star phase, $\Delta t_{\rm [X_{\rm 1},X_{\rm 2}],m}$ is 
the amount of time the value of the observable $X_{\rm obs}$ lies 
between $X_{\rm 1}$ and $X_{\rm 2}$ for a given model m, $\Delta t_{\rm BH+O,m}$ 
is the total duration of the BH+OB star phase of the model 
m. Lastly, $W_{\rm m}$ is the birth weight of each model given by
\begin{equation}
    W_{\rm m} = \rm{log}\,(M_{\rm 1,i}/M_{\odot})^{-1.35} \cdot q_{\rm i}^{-1.0} \cdot \rm{log}\,(P_{\rm i}/\rm{d})^{-0.45}.
    \label{equation_weight_factor}
\end{equation}

Using the above definitions, we derive the distributions of 
observable properties during the BH+OB phase of our models. 
As such, the histograms show the distribution of observables in 
an unbiased, ideal and complete sample of BH+OB star binaries. 
Finally, to compare the X-ray luminosity arising from shocks in the 
OB star wind ($L_{\rm X,wind}$) to the X-ray luminosity arising 
from the vicinity of the BH, we take $L_{\rm X,wind}$ as $10^{-7}$ 
times the bolometric luminosity of the OB star 
\citep{Feldmeier1997,Huenemoerder2012,Crowther2022,Bernini-Peron2023}. 


\section{Results}
\label{section_results}

\subsection{Mass accretion rate}
\label{section_mass_accretion_rate}

Figure\,\ref{figure_Macc_Mnet} shows the distribution of 
the mass accretion rate at the accretion radius (left panel) 
and the net mass accretion rate near the event horizon of the 
BH (right panel). Higher mass accretion rates are reached for 
shorter orbital period systems 
(Eq.\,(\ref{equation_mass_accretion_rate}), expressed in terms 
of orbital separation). The peak in the mass accretion rate 
distribution is $\sim10^{-11}-10^{-10}\,M_{\odot}\, 
\mathrm{yr}^{-1}$ and $\sim10^{-12}-10^{-11}\,M_{\odot}\, 
\mathrm{yr}^{-1}$ for 1.4-10\,d and 10-100\,d initial orbital 
periods respectively. The shortest-period systems can also form 
an accretion disk around the BH \citep{Sen2021}, denoted by 
the black hatching. We note however that there is a significant 
contribution from the 10-100\,d binaries to the population 
of BH+OB star systems that can form an accretion disk (see 
Sect.\,\ref{secttion_X-ray-bright}). 

The highest mass accretion rates ($\geq\,10^{-8}\,M_{\odot}\, 
\mathrm{yr}^{-1}$) is mostly from models in the 10-100\,d initial 
orbital period range. The shortest initial orbital period models 
($<2$\,d) enter a contact phase and merge on the main sequence 
\citep{Menon2021}. Many of the short-period binary models 
(2-10\,d) in our grid do not survive their prior Case\,A and 
Case\,AB mass transfer phase (\citealp[Fig.\,1 of][]{Sen2022}, 
\citealp[Appendix\,C of][]{Pauli2022}) to reach the BH+OB star 
phase. 

The total number of short-period (1.4-10\,d) binary models 
contributing to the distribution of mass accretion rate is 
much smaller than the number of models with longer initial 
orbital periods (10-100\,d). Correspondingly, the peak in 
the distribution of mass accretion rate does not occur at 
the highest mass accretion rates but at lower mass accretion 
rates where more models contribute to the histogram function 
(Eq.\,(\ref{equation_histogram})), despite our assumed intrinsic
period distribution favouring short-period binaries 
(Eq.\,(\ref{equation_weight_factor})). 

The mass accretion rate typically decreases for models with longer 
orbital periods (Eq.\,(\ref{equation_mass_accretion_rate})). The 
drop in the number of systems with mass accretion rates below 
$10^{-8}\,M_{\odot}\,\mathrm{yr}^{-1}$ is produced by our 
assumption that longer-period binaries are less likely to be born 
(Eq.\,(\ref{equation_weight_factor})). Hence, the peak in the 
distribution of mass accretion rates arises at $\sim$10$^{-12}$-10$^{-11}$\,$M_{\odot}$$\,$yr$^{-1}$, 
from BH+OB star binaries with intermediate orbital periods. 

We assume that mass outflows reduce the mass accretion 
rate from the accretion radius to the BH event horizon 
(Eq.\,\ref{equation_net_mass_accretion_rate}). The right 
panel shows that the peak in the distribution of the net 
mass accretion rate near the event horizon of the BH is 
$\sim$2 orders of magnitude lower than that at the accretion 
radius. This reduction in mass accretion rate results in a 
proportionate decrease in X-ray luminosity (Eq.\,\ref{equation_Lx}) 
predicted from ADAFs around BHs in the BH+OB star binaries. 
We note that observations have confirmed the presence of 
outflows in Sgr\,A* \citep{Hawley2002,Yuan2003,Igumenshchev2003}. 
In case the same may not true for accretion onto stellar-mass 
BHs (e.g., see \citealp{Fender2014}), our predictions for 
the X-ray luminosity will be $\sim$\,2 orders of magnitude 
higher than presented in the next section. 

\begin{figure*}
    \includegraphics[width=0.48\hsize]{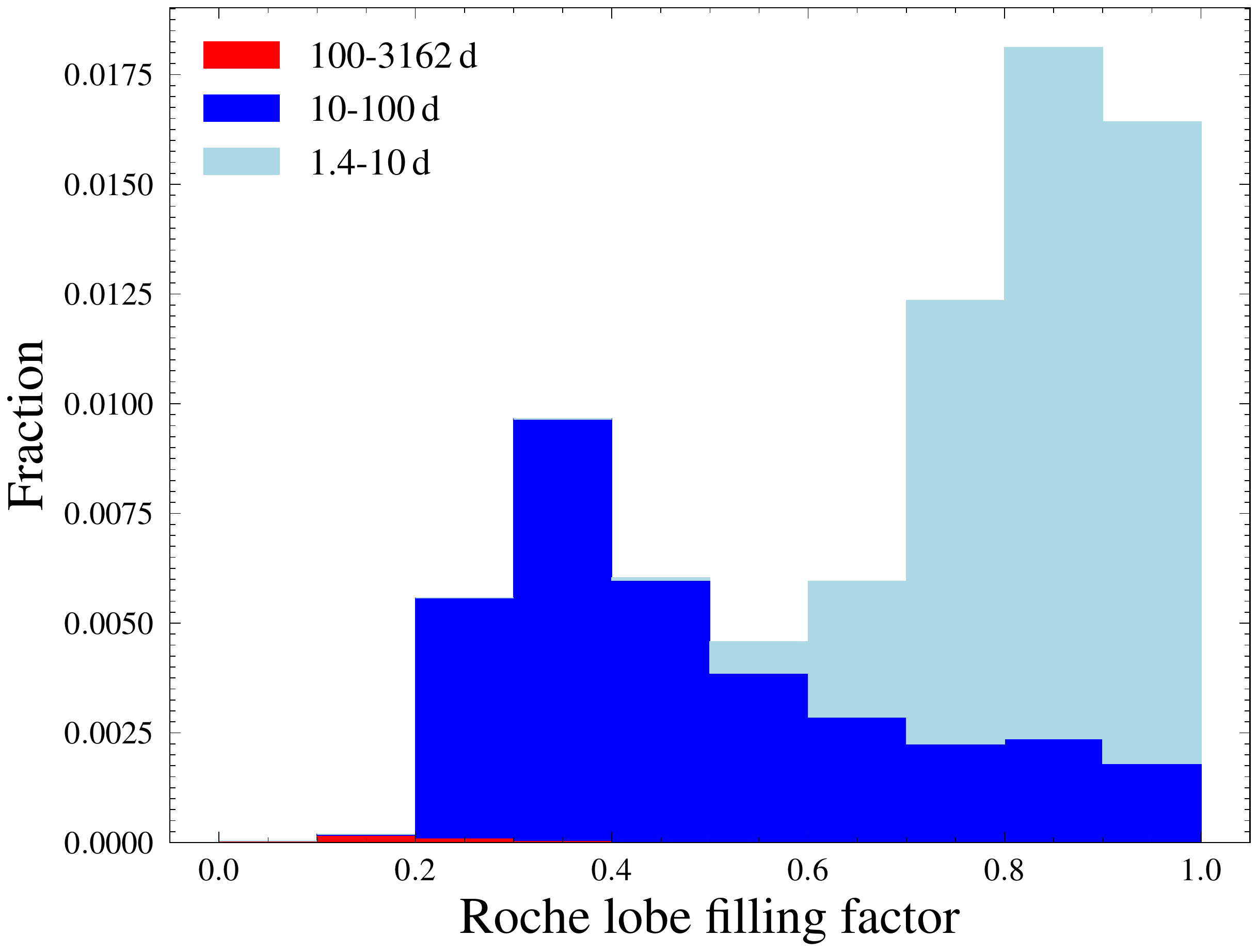}
    \includegraphics[width=0.48\hsize]{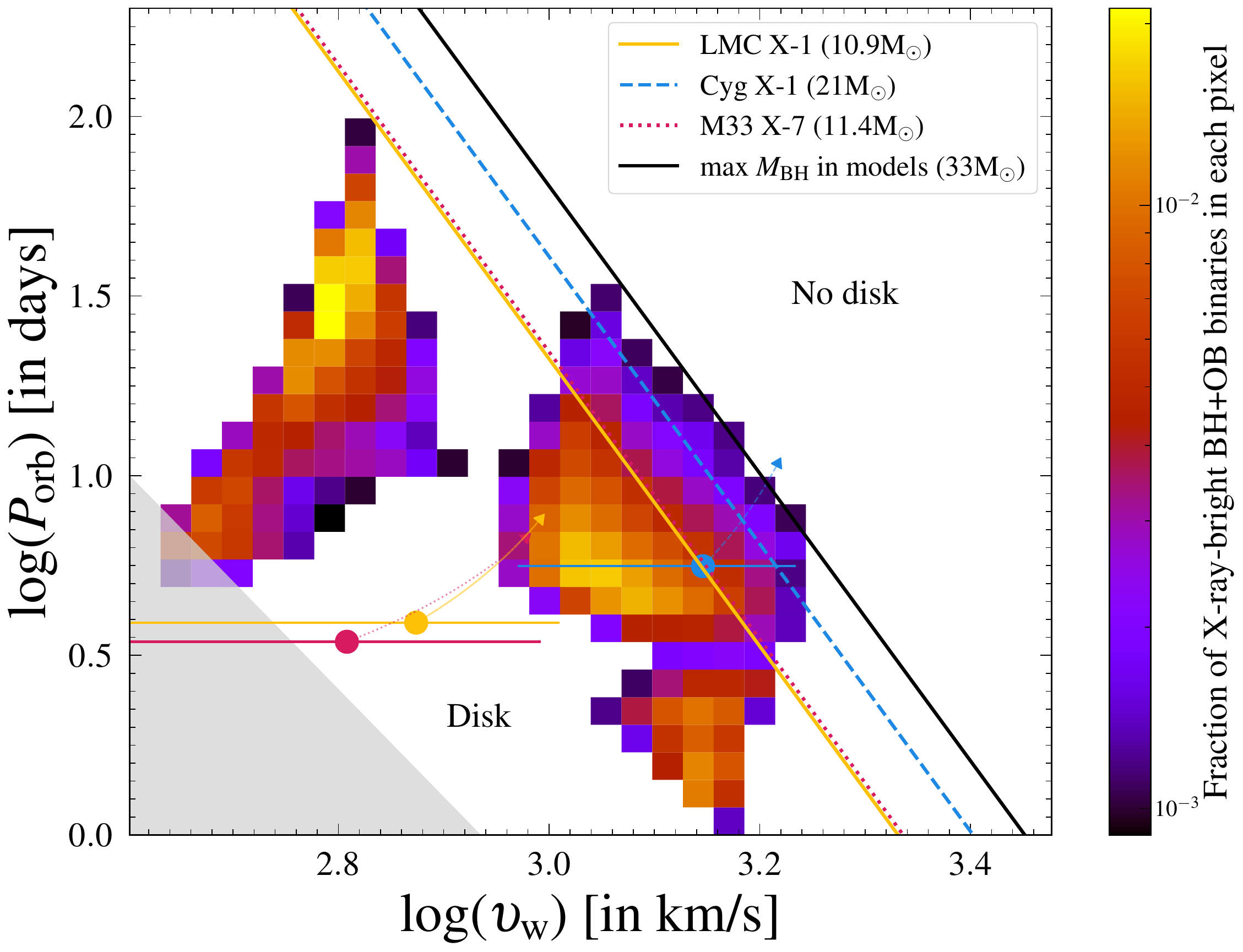}
    \caption{\textit{Left panel}: Distribution of the Roche lobe filling factor of BH+OB star binaries where an accretion disk can form (Sect.\,\ref{sec:disk_formation}). Light blue, dark blue and red colours denote the contributions from binary models having different initial orbital period ranges of 1.4-10\,d, 10-100\,d and 100-3162\,d respectively. The coloured histograms are stacked on each other, and the sum of their ordinate values equals 0.0785 (the fraction of X-ray-bright systems predicted from our entire population of BH+OB star binaries). \textit{Right panel}: Disk formation thresholds (equation (10) of \citealp{Sen2021}) shown by slanted solid, dashed and dotted lines for three BH masses corresponding to the three observed wind-fed BH HMXBs \citep{Orosz2007,Orosz2009,Miller-Jones2021,Ramachandran2022}. The black slanted line shows the disk formation threshold of the maximum BH mass from our binary model grid. An accretion disk can form below the slanted lines and vice versa. The three circles correspond to the observed orbital period and calculated OB star wind speed at the position of the BH (assuming a $\beta$-law for the wind velocity profile, with $\beta$ = 1, equation\,1 of \citealp[]{Sen2021}) of the three BH-hosting wind-fed HMXBs. The horizontal solid lines correspond to the range of wind speed by varying $\beta$ from 0.5 (upper-speed limit) to 2 (lower limit). The arrows indicate the new position of the square markers if the BH's orbital period was twice as long. The grey region is where the wind accretion approximation $\upsilon_{\rm w}>a\Omega$ is no longer valid. The heatmap shows the distribution of the orbital period of the binary models and wind speed of the OB star companions during the BH+OB star phase when an accretion disk can form. The colour bar gives the predicted fraction of BH HMXBs in each pixel. The sum of the pixels is equal to unity. }
    \label{figure_X-ray-bright}
    \label{fig:disk_form}
\end{figure*}

\subsection{X-ray luminosity}

Figure\,\ref{figure_Lx} shows the X-ray luminosity distribution 
from BH+OB star binaries according to our binary evolution models 
due to accretion onto the BH (coloured histograms) and shocks 
in the wind of the OB star companion (black step histogram). In 
all the panels, the X-ray luminosity in the range $10^{\rm 35} 
- 10^{\rm 38}$\,erg\,s$^{-1}$ originate solely from a Keplerian 
accretion disk \citep{Shakura1973} around the BHs (Eq.\,\ref{Lx_eqn}). 
These strong X-ray sources comprise 7.85\% of the entire population 
of BH+OB star binaries in our grid. The orbital period distribution 
(see figure 6 of \citealp{Langer2020}) of the BH+OB star binaries 
peaks above $\sim$100\,d, where the strong wind velocity of the 
OB star companion disfavours the formation of an accretion disk 
\citep{Sen2021}. Hence, most BH+OB star binaries in our model grid 
do not form an accretion disk. We call the BH+OB star binaries with 
X-ray luminosity above $10^{\rm 35}$\,erg\,s$^{-1}$ `X-ray-bright'. 

We find that the contribution from binary models in the 40-90\,$M_{\odot}$ 
range to the population of X-ray-bright BH+OB star binaries is negligible. 
We identify two reasons. Firstly, the IMF strongly disfavours the most 
massive binaries in the population. Secondly, the most massive models 
enter a contact phase up to an initial orbital period of $\sim$3\,d 
\citep[figures\,C1-C3 of][]{Pauli2022}. Hence, the shortest-period models, 
likely to form an accretion disk during a BH+OB phase, do not 
reach their BH+OB star phase. 

The X-ray luminosity distribution shows a second, broader peak at 
luminosities in the range from $10^{\rm 25} - 10^{\rm 35}$\,erg\,s$^{-1}$ 
which is associated with systems having ADAFs \citep{Narayan1995} around 
BHs that do not have Keplerian accretion disk around them 
(Sect.\,\ref{section_acc_disk_absent}). The largest contribution 
comes from the binaries with initial primary masses of 20-40\,$M_{\odot}$. 
A large fraction of the primaries in the initial mass range of 
10-20\,$M_{\odot}$ have helium core mass smaller than 6.6\,$M_{\odot}$ 
such that they are not expected to collapse into BHs while 
the contribution from the 40-90\,$M_{\odot}$ range is suppressed 
by the IMF and comparatively shorter lifetimes of more massive 
O-star companions. 

For the case of the most inefficient viscous coupling $\delta$ = 
0.001 (top panel), our models predict that 20.55\% of the BH+OB 
star binaries in the LMC to have X-ray luminosities between 
$10^{\rm 31}$-$10^{\rm 33}$\,erg\,s$^{-1}$. We take the lower 
cut-off of observable X-ray luminosity to be $10^{\rm 31}$\,erg\,s$^{-1}$ 
based on X-ray detection from O and B stars in the LMC with the 
\textit{Chandra} Visionary programme T-ReX \citep[see figure\,3 
of][]{Crowther2022}. Our results imply that for the one X-ray-bright 
BH+OB binary (LMC X-1, \citealp{Orosz2009}) found in the LMC, we 
expect $\sim$2.6 faint X-ray sources observable with \textit{Chandra}, 
for the least efficient viscous heating (see also, 
Sect.\,\ref{section_numbers}). We call the BH+OB star binaries 
with X-ray luminosity between 
$10^{\rm 31}$-$10^{\rm 35}$\,erg\,s$^{-1}$ `X-ray-faint'. 

\citet[][figure\,1]{Xie2012} found that the radiative efficiency increases 
by $\sim$one order of magnitude each time when the efficiency 
of viscous coupling is increased from $\delta$ = 0.001 to 0.1 
to 0.5. This is reflected in the distribution of X-ray luminosity 
from the middle and bottom panels in the $10^{\rm 25} - 10^{\rm 35}$\,erg\,s$^{-1}$ 
range. The maximum value of X-ray luminosity reaches 10$^{34}$\,erg\,s$^{-1}$ 
and 10$^{35}$\,erg\,s$^{-1}$ for $\delta$ = 0.1 and 0.5, respectively. 
This increases the number of observable BH+OB star binaries in X-rays. 
For efficient viscous heating parameters of $\delta = 0.1$ and 
$\delta = 0.5$, our models predict $\sim$4.1 and $\sim$6.8 X-ray-faint 
BH+OB star binaries in the LMC, respectively. 

The bolometric luminosity of the OB star companions during the 
BH+OB star phase is in the range $\rm{log}\,L/L_{\odot}=
4\dots6$. 
The X-ray luminosity that arises from shocks embedded in the 
wind of the OB star companion (L$_{\rm X,wind}$) ranges from 
$10^{\rm 30} - 10^{\rm 33}$\,erg\,s$^{-1}$ in our models. 
Hence, the wind X-ray luminosity from the OB star is in the 
same range as our predictions of the X-ray luminosity from 
advection-dominated accretion flows around the BH. 
However, the X-ray emission originating from shocks in the 
OB star wind is thermalised, while the X-ray emission from 
the surrounding of the BH will be non-thermal in the case 
of advection-dominated accretion \citep{Ichimaru1977,Narayan1998}. 
We discuss the possibility of disentangling the two contributions 
to the total X-ray luminosity in Sect.\,\ref{section_disentangling}. 

The X-ray luminosity of observed Be X-ray binaries is higher 
than 10$^{34}$\,erg s$^{-1}$, even for the faint persistent 
sources \citep{Cheng2014}. For inefficient viscous heating 
of electrons, the X-ray luminosity predicted from the BH+O 
star binaries remain below the lower limit of the 
X-ray luminosity from Be X-ray binaries (top and middle 
panels). For $\delta$ = 0.5 (bottom panel), only 2\% of the 
BH+O star binaries may have X-ray luminosity between 
$10^{\rm 34} - 10^{\rm 35}$\,erg\,s$^{-1}$, which is lower 
than the fraction of BH+O star binaries predicted 
to have a Keplerian accretion disk around the BH. 
We discuss this in more detail in Sect.\,\ref{section_be_xray_binaries}. 

\begin{figure*}
    \centering
    \includegraphics[width=0.48\hsize]{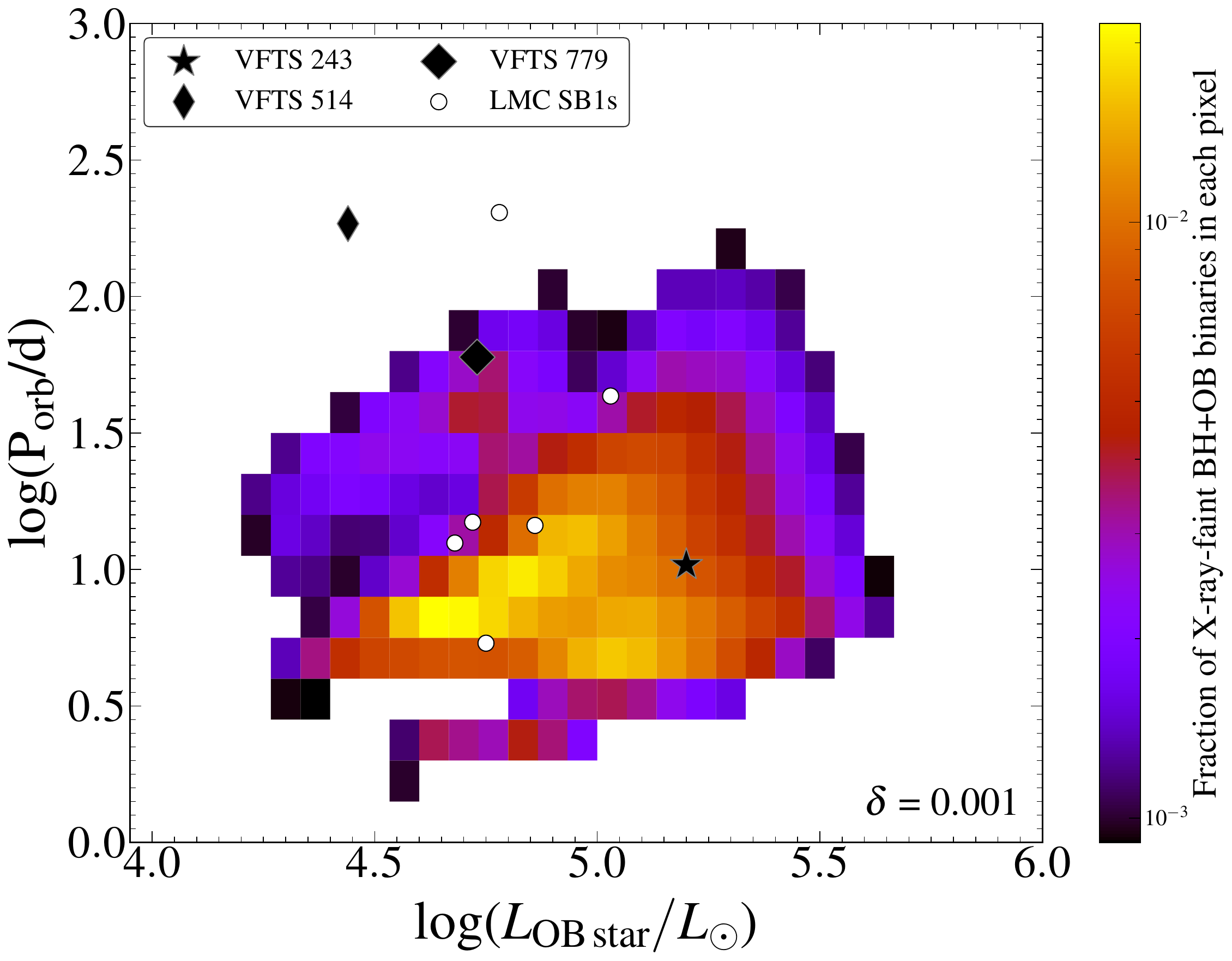}
    \includegraphics[width=0.48\hsize]{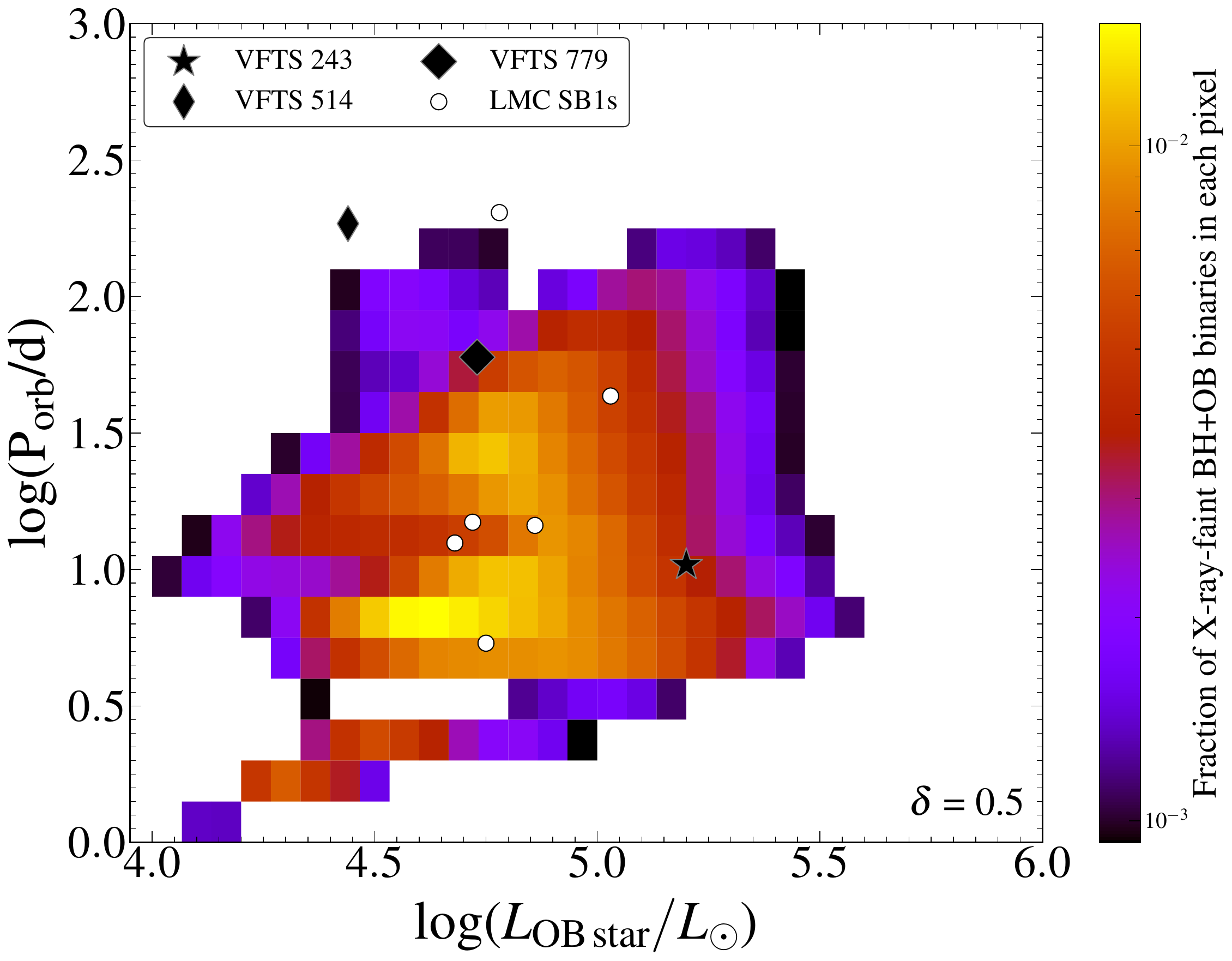}
    \caption{Distribution of orbital period and luminosity of the OB star companion from BH+OB star binaries that can produce X-ray luminosity 10$^{31}$-10$^{35}$\,erg/s through wind accretion, for two values of the viscous heating parameter $\delta$ = 0.001 (\textit{left panel}) and 0.5 (\textit{right panel}). The colour bar gives the predicted fraction of X-ray-faint BH+OB star binaries in each pixel. The sum of the pixels in each panel equals unity. VFTS\,243 \citep{Shenar2022} is shown by a black star. Two more potential BH+OB star binaries are marked with thin and thick diamonds \citep{Shenar2022}. Six more SB1 systems with unseen companions (having mass above 2.35\,$M_{\odot}$) are marked by white circles (Table\,\ref{table_observations}; data from \citealp{Shenar2022}). }
    \label{figure_Porb_Lo}
\end{figure*}

\subsection{X-ray-bright systems}
\label{secttion_X-ray-bright}

The left panel of Fig.\,\ref{fig:disk_form} shows the distribution 
of the Roche lobe filling factor of the OB star companion when an 
accretion disk can form during the BH+OB star phase of the binary 
models. We find a distinct bimodal distribution, partly as a natural 
consequence of the binary orbital periods (1.4\,d-10d vs 10-100\,d) 
that contribute to the different peaks (0.3 vs 0.9). The filling factors 
of models arising from 1.4-10\,d models peak at $\sim$0.9. The 
orbital periods are short enough that the accretion disk can form 
only towards the end of the main sequence of the O star companions 
when the stellar radius increases and the wind speed decreases. 
On the other hand, the contribution from 10-100\,d binary models 
comes primarily from BHs with B star companions, where the terminal 
velocity of the B star wind is a factor of two lower than O stars. 
Moreover, many binary models with small initial mass ratios merge 
on the main sequence at the 1.4-10\,d orbital period range 
\citep[e.g., figure\,1 and figure\,F1 of][]{Sen2022}. 

The right panel of Fig.\,\ref{fig:disk_form} shows the distribution 
of the X-ray-bright BH+OB models on the orbital period-wind 
speed plane, where they occupy two separate regions. The binaries 
with orbital periods of 1-20\,d during the BH+OB star phase have 
companions that have wind speeds above 1000\,km$\,$s$^{-1}$. 
The companions in these binaries are mostly O stars originating 
from initial orbital periods of 1.4-10\,d, and have high Roche-lobe 
filling factors (c.f. left panel). On the contrary, the peak at 
orbital period $\sim$30\,d and wind speed $\sim$600\,km$\,$s$^{-1}$ 
comes from BHs with less massive, B-star companions. These models 
have initial orbital periods $\sim$10-100\,d and initial mass 
ratios between 0.6-0.3. We also note the absence of predicted 
wind-fed HMXBs at short orbital periods and low wind speeds, 
because our short-period models with low initial mass ratios 
generally merge on the main sequence. 

\subsection{X-ray-faint systems}
\label{section_X-ray-faint}

Figure\,\ref{figure_Porb_Lo} shows the orbital period and OB 
star luminosity of the BH+OB star binaries that are predicted 
to emit X-ray emission above 10$^{31}$\,erg s$^{-1}$, the 
observational threshold of current X-ray telescopes/surveys 
\citep{Crowther2022}. For $\delta$ = 0.001, the predicted 
distribution of orbital periods peak near $\sim$10\,d. The 
mass accretion rate falls inversely with the square of the 
orbital separation (Eq.\,\ref{equation_mass_accretion_rate}), 
thereby making higher orbital period binaries less likely to 
overcome the X-ray luminosity threshold in the LMC. At the 
shortest orbital periods, all our binary models do not survive 
the mass transfer phases before the BH+OB star phase 
(Sect.\,\ref{section_mass_accretion_rate}), due to which the 
probability distribution decreases below 10\,d. 

The wind mass loss rate of the OB stars increases with 
luminosity (see, e.g., \citealp{Kudritzki2000,Puls2008}) 
and mass \citep{Vink2000}. BHs in orbit with more massive 
(luminous) stars at the same orbital period have a greater 
mass accretion rate than less massive (luminous) stars 
(Eq.\,\ref{equation_mass_accretion_rate}), as the mass 
accretion rate is proportional to the wind mass-loss rate. 
Consequently, the luminosity of the OB star has to be 
higher at long orbital periods to keep the mass accretion 
rate onto the BH high enough to produce observable X-ray 
emission. As such, we see that the distribution of observable 
BH+OB star binaries extends towards higher luminosities 
at higher orbital periods. 

For $\delta$ = 0.5 (right panel of Fig.\,\ref{figure_Porb_Lo}), 
the radiative efficiency $\epsilon$ increases by $\sim$\,2 
orders of magnitude compared to $\delta$ = 0.001, which leads 
to higher X-ray luminosity produced from the same mass accretion 
rate. We see a larger contribution from longer-period binaries 
to the distribution of BH+OB star binaries with observable X-ray 
emission. The increased radiative efficiency compensates for the 
lower mass accretion rate at longer periods. For high efficiencies 
of viscous dissipation, a significant number of our BH+OB star 
models with orbital periods $\sim$100\,d can produce observable 
X-rays in the LMC (see also Fig.\,\ref{figure_Porb_Lo_appendix}, 
for $\delta$ = 0.1). 

\subsection{Absolute number of X-ray-faint BH+OB star binaries in the LMC}
\label{section_numbers}

\citet{Langer2020} studied the distribution of BH+OB star 
binaries arising from the initial primary mass range of 
10-40\,$M_{\odot}$. Assuming constant star formation, they 
estimated $\sim$120 BH+OB star binaries from the above 
parameter space in the LMC. The relative contributions 
from the 10-40\,$M_{\odot}$ and 40-90\,$M_{\odot}$ range 
to the total population of BH+OB star binaries in our study 
is $\sim$88\% and $\sim$12\%, respectively (Fig.\,\ref{figure_Lx}). 
This implies an additional $\sim$16 BH+OB star binaries 
from the 40-90\,$M_{\odot}$ range of initial primary masses. 

Our models predict that 7.85\% of the total population 
of BH+OB star binaries from the 10-90\,$M_{\odot}$ are 
have X-ray luminosity above 10$^{35}$\,erg\,s$^{-1}$ 
(Fig.\,\ref{figure_Lx}). 
This implies that our models predict $\sim$10 X-ray-bright 
BH+OB star binaries in the LMC. Yet, only one wind-fed BH 
HMXB has been observed in the LMC (LMC X-1, \citealp{Orosz2009}). 
We discuss this discrepancy in the number of observed 
to predicted wind-fed BH HMXBs in Sect.\,\ref{section_accretion_disk_formation}.

Of the $\sim$126 BH+OB star binaries that are not 
expected to form an accretion disk around the BH, 
the number of X-ray-faint BH+OB star binaries depend 
on the efficiency of viscous heating adopted in our 
models. The predicted numbers range among 28\dots44\dots72 
for $\delta$ = 0.001\dots0.1\dots0.5, respectively. 
Thus, a significant handful of BH+OB star binaries in 
the LMC are expected to produce faint yet observable 
X-ray emission, even for the most inefficient case 
of viscous heating. Moreover, if accretion disk formation 
is inhibited in some of the BH+OB binaries that are 
predicted to be X-ray-bright in this work, the high 
mass accretion rates predicted in the X-ray-bright 
systems will make them observable as X-ray-faint 
sources. Hence, the number of X-ray-faint BH+OB star 
binaries will increase for more stringent constraints 
on the accretion disk formation criterion 
(Sect.\,\ref{section_accretion_disk_formation}).


\section{Comparison with observations}
\label{section_observations}

Table\,\ref{table_observations} lists confirmed and tentative BH+OB 
star binaries observed in the LMC. LMC X-1 is a long-studied 
system with continuous X-ray emission produced from a Keplerian 
accretion disk (see, e.g. \citealp{Pakull1986,Orosz2009}). VFTS\,243 
is a recently discovered X-ray inactive BH+O star binary in 
the Tarantula Nebula \citep{Shenar2022n}. 
\citet{Shenar2022} identified two more systems, VFTS 514 and 779, as 
strong candidates to host BHs that do not show strong X-ray emission. 
The remaining systems are SB1 binaries with unseen companions of mass 
above 2.35\,$M_{\odot}$\footnote{The most massive neutron star discovered is 
$\sim$2.35\,$M_{\odot}$ \citep{Romani2022}} \citep[Table\,2 of][]{Shenar2022}. 

\begin{table}
\caption{Confirmed and plausible BH+OB star binaries in the LMC.}             
\label{table_observations}      
\centering                          
\begin{tabular}{c c c c}        
\hline\hline                 
Name & Period [d] & log($L_{\rm OB\,star}/L_{\odot}$) & log($L_{\rm x}/\rm{erg\,s^{-1}}$) \\    
\hline                        
LMC X-1  & 3.90   & 5.33   & 38.28  \\      
\hline
VFTS 243 & 10.40  & 5.20   & <32.15 \\      
\hline
VFTS 514 & 184.92 & 4.44   & $--$   \\
VFTS 779 & 59.94  & 4.73   & $--$   \\
\hline
VFTS 619 & 14.5   & 4.86   & $--$   \\
VFTS 631 & 5.37   & 4.75   & $--$   \\
VFTS 645 & 12.5   & 4.68   & $--$   \\
VFTS 743 & 14.9   & 4.72   & $--$   \\
VFTS 827 & 43.2   & 5.03   & <32.48 \\ 
VFTS 829 & 203.0  & 4.78   & $--$   \\
\hline                                   
\end{tabular}
\tablefoot{The data for LMC X-1 are taken from \citet{Orosz2009} and \citet{Pakull1986}. The data for the VFTS systems are taken from \citet{Shenar2022}.}
\end{table}

\subsection{X-ray-bright systems}

In the right panel of Fig.\,\ref{fig:disk_form}, we use the criterion 
for the formation of an accretion disk to show the domains where a disk 
is expected or not (with $\bar{s}=0$ and $\eta=1/3$), for black hole 
masses corresponding to three high-mass X-ray binaries: LMC X-1 
\citep{Orosz2009}, Cygnus X-1 \citep{Orosz2011} and M33 X-7 
\citep{Orosz2007,Valsecchi2010,Ramachandran2022}. The wind speed 
of the main sequence 
companion to the BH is computed at the orbital separation with a 
$\beta$-law \citep[equation 1 of][]{Sen2021}. Each coloured 
circle marker lies below its respective threshold for accretion disk 
formation. This shows that even if the OB star companion does not fill 
its Roche lobe, an accretion disk can form through wind capture in 
these three systems, from where copious amounts of X-rays can be 
emitted. In the grey region, the wind launching is quenched and 
mass accretion onto the BH can only proceed through Roche lobe 
overflow. 

We see that the binary models cannot reproduce the position of 
LMC\,X-1 in the orbital period-wind speed plane. The mass of the 
BH and the O star companion is 10.91$\pm$1.41\,$M_{\odot}$ and 
31.79$\pm$3.48\,$M_{\odot}$ respectively. However, our models 
at the relevant initial mass (30-50\,$M_{\odot}$, that can form 
a $\sim$10\,$M_{\odot}$ BH), and initial orbital period (1-3\,d) 
range merge on the main sequence after a contact phase. This may 
indicate the short period may detach after a brief contact phase. 
Investigating the physics of the contact phase is beyond the scope of 
this work \citep[see, e.g.][]{Fabry2022,Fabry2023}. Alternatively, 
at an orbital period of 3.9\,d the black hole in LMC\,X-1 is deeply 
engulfed in the wind acceleration region of the 31.8\,$M_{\odot}$ 
O star. X-rays from the BH can ionize the stellar wind decreasing 
its speed \citep{Krticka2009}. This effect is not taken into account 
in our simple analysis. We assume that the O star companion has 
a radiatively driven wind which follows the $\beta$-law \citep[as 
in ][]{Orosz2007}.

On the other hand, our models predict a sub-population of X-ray-bright 
BH+OB star binaries above 15\,d with wind speeds below 800\,km$\,
\rm{s}^{-1}$. The OB star companions in this population also have a low 
filling factor. There are no observed wind-fed HMXBs in this region. 
Despite the low-number statistics, this may imply two possibilities. 
First, binaries may not undergo stable mass transfer at low initial 
mass ratios and long orbital periods \citep[see however,][where the 
criterion for unstable mass transfer is more relaxed]{Schurmann2024}. 
On the other hand, our criterion to determine the formation of an 
accretion disk may be inadequate to filter out BH+OB models where 
the OB star has low filling factors. We discuss this further in 
Sect.\,\ref{section_accretion_disk_formation}. 

\subsection{X-ray-faint systems}

\subsubsection{HD96670}

HD96670 is a single-line spectroscopic binary in the Carina OB2 
association, tentatively hosting a BH of mass 6.2\,$M_{\odot}$ 
in orbit with an O star of mass $\sim$22.7\,$M_{\odot}$ and 
radius $\sim$\,17.1\,$R_{\odot}$. The orbital period of the binary 
is 5.28\,d \citep{Gomez2021}. The orbital separation of the binary 
is $\sim$39.2\,$R_{\odot}$, and the Roche-lobe filling factor 
of the O star is $\sim$0.885. Assuming the effective temperature 
of the O star to be 38000\,K \citep{Hohle2010}, we estimate its 
luminosity to be log($L/L_{\odot}$)$\sim$5.75. 

We assume the ratio of the terminal velocity to the escape velocity 
for O stars to be 2.6 \citep{Vink2000,Kudritzki2000}, $\beta$=1 in 
the wind velocity law, and the Eddington factor at the surface of the 
O star to be $\sim$0.2 (see equations 1-8 of \citealp[]{Sen2021}). 
Then, the estimated wind velocity $\upsilon_{\rm w}$ of the O star at 
the position of the BH to be $\sim$932.3\,km\,s$^{-1}$. The orbital 
velocity is $\sim$375.4\,km\,s$^{-1}$. For the fiducial value of 
accretion efficiency ($\eta$=1/3) and a non-spinning BH, we find 
that an accretion disk does not form (equation\,10 of \citealp{Sen2021}). 
This is consistent with the lack of bright X-ray emission from 
this system.

We estimate the wind mass-loss rate to be one-third of the mass-loss 
rate derived from equation\,24 of \citet{Vink2001}, $\dot{M}_{\rm w}$
$\sim$10$^{-5.71}$\,$M_{\odot}$\,yr$^{-1}$. The accretion radius, 
mass accretion rate at the accretion radius $\dot{M}_{\rm acc}$, net 
mass accretion rate near the BH event horizon $\dot{M}_{\rm net}$ 
and Eddington mass accretion rate of the BH $\dot{M}_{\rm Edd}$ can 
be estimated to be 2.36\,$R_{\odot}$, 
2.85$\times$10$^{-9}$\,$M_{\odot}$\,yr$^{-1}$, 
4.85$\times$10$^{-11}$\,$M_{\odot}$\,yr$^{-1}$ and 
1.33$\times$10$^{-7}$\,$M_{\odot}$\,yr$^{-1}$, respectively. 
For $\dot{M}_{\rm net}$/$\dot{M}_{\rm Edd}$ = 0.00036, the 
radiative efficiency is $\sim$0.003 (figure\,1 of \citealp{Xie2012}). 
The above quantities lead to an estimated X-ray luminosity of 
8$\times$10$^{33}$\,erg\,s$^{-1}$ from ADAF. The observed X-ray 
flux from this system ranges from 2.2$\times$10$^{32}$\,erg\,s$^{-1}$ 
(NuSTAR, \citealp{Gomez2021}) to 2.4$\times$10$^{34}$\,erg\,s$^{-1}$ 
(XMM-Newton, \citealp{Saxton2008}). Our results are in the right 
ballpark, although further study may be required to accurately 
constrain the X-ray variability and/or photometric properties of 
the system \citep[e.g., see][]{HanyueWang2022}. 

\subsubsection{VFTS 243}

The constrained stellar and binary parameters of VFTS\,243 enable 
us to estimate the X-ray luminosity from ADAF around its 
$\sim$10\,$M_{\odot}$ BH. The mass, radius and luminosity of the 
O star companion is $\sim$25\,$M_{\odot}$, $\sim$10.3\,$R_{\odot}$ 
and $\sim 10^{5.2}\,L_{\odot}$ 
respectively. The wind mass loss rate, with a clumping factor of 10, 
is $\sim$1.5$\times10^{-7}$\,$M_{\odot}\,$yr$^{-1}$. For the adopted 
terminal wind velocity of 2100\,km$\,$s$^{-1}$, we estimate the 
net mass accretion rate near the event horizon $\sim$1.6$\times10^{-13}$\,$M_{\odot}\,$yr$^{-1}$. 
The Eddington mass accretion rate of a non-spinning 10\,$M_{\odot}$ 
BH is $\sim$2.2$\times10^{-7}$\,$M_{\odot}\,$yr$^{-1}$, leading to a 
ratio $\dot{M}_{\rm net}/\dot{M}_{\rm edd}$$\sim$7.5$\times10^{-7}$. 
The radiative efficiency of an ADAF at the above ratio of 
$\dot{M}_{\rm net}/\dot{M}_{\rm edd}$ is $\sim$0.0001, $\sim$0.001 
and $\sim$0.01 \citep[extrapolated from Fig.\,1 of][]{Xie2012} for 
$\delta$ = 0.001, 0.1 and 0.5 respectively. The resulting X-ray 
luminosity is $\sim$10$^{29}$, $10^{30}$ and $10^{31}$\,erg$\,$s$^{-1}$ 
respectively. Hence, VFTS\,243 is not expected to be observable 
even in faint X-rays. The \textit{Chandra} T-ReX programme 
\citep{Crowther2022} constrained the X-ray luminosity of this system
log($L_{\rm X}$/$\rm erg\,s^{-1}$) < 32.15, which is within our 
above estimates. 

We note however that even though VFTS\,243 may lie within the 
parameter space of Fig.\,\ref{figure_Porb_Lo}, our detailed 
calculation above shows it will not be observable in faint X-rays. 
This is because the parameter space for BH+OB star binaries showing 
observable faint X-ray emission is degenerate with the parameter 
space for BH+OB star binaries that do not show faint X-ray emission 
(c.f. Fig.\,\ref{figure_Porb_Lo} and Fig.\,6 of \citealp[]{Langer2020}). 
We note however that the orbital period distribution of all BH+OB 
star binaries peak above $\sim$100\,d, while the ones that may be 
identifiable via faint X-rays peak around $\sim$10\,d. 

\subsubsection{VFTS 399}

The VLT Flames survey in the Tarantula nebula \citep{Evans2011} 
identified an SB1 system, VFTS\,399, which has a O9 IIIn star 
with a compact object companion \citep{Clark2015}. \citet{Clark2015} 
reported an X-ray luminosity of $5\times10^{34}$\,erg\,s$^{-1}$ 
while an orbital solution could not be obtained. An orbital 
solution was neither presented in the recent work of \citet{Villasenor2021} 
for this object. The system also has a very low dispersion 
velocity $\sigma$\,$\approx$\,10\,km\,s$^{-1}$ relative to its environment 
\citep{sana2013}. \citet{Clark2015} also note that for the 
empirical lower bound on the orbital period distribution of 
$\sim$20\,d of Be X-ray binaries \citep{Cheng2014,Haberl2022}, 
the mass of the compact object companion in VFTS\,399 has to be greater 
than 2.5\,$M_{\odot}$, based on their radial velocity measurements. 
For higher orbital periods, the mass of the compact object 
companion is still higher, hinting at the possibility of a BH, 
although apparently excluded by the detection of X-ray pulsations. 

\subsubsection{Other systems}

Figure\,\ref{figure_Porb_Lo} shows that both for $\delta$ = 0.001 
(left panel) and $\delta$ = 0.5 (right panel), the orbital period 
of VFTS 514 is longer than the predicted distribution of X-ray-faint 
BH+OB star binaries from our grid. The position of VFTS\,779 
shows that it may be observable in faint Xrays for $\delta$= 0.5. 
Five of the six SB1 systems 
are near the peak of the predicted distribution of X-ray-faint 
BH+OB binaries. We note that there is a lack of X-ray observations 
of SB1 systems having B-type stars with log($L/L_{\odot}$) $\lesssim$ 
5 \citep{Crowther2022}. Unambiguous identification of the 
presence of a BH in these systems and a targeted X-ray observation 
programme on these systems may provide empirical evidence to constrain 
the strength of viscous coupling in hot accretion flows around 
stellar-mass BH binaries. For example, \citet{Villasenor2021} 
identified 16 SB1 systems with B-type stars having a high probability 
of hosting compact object companions. 

\begin{figure*}
    \centering
    \includegraphics[width=0.48\hsize]{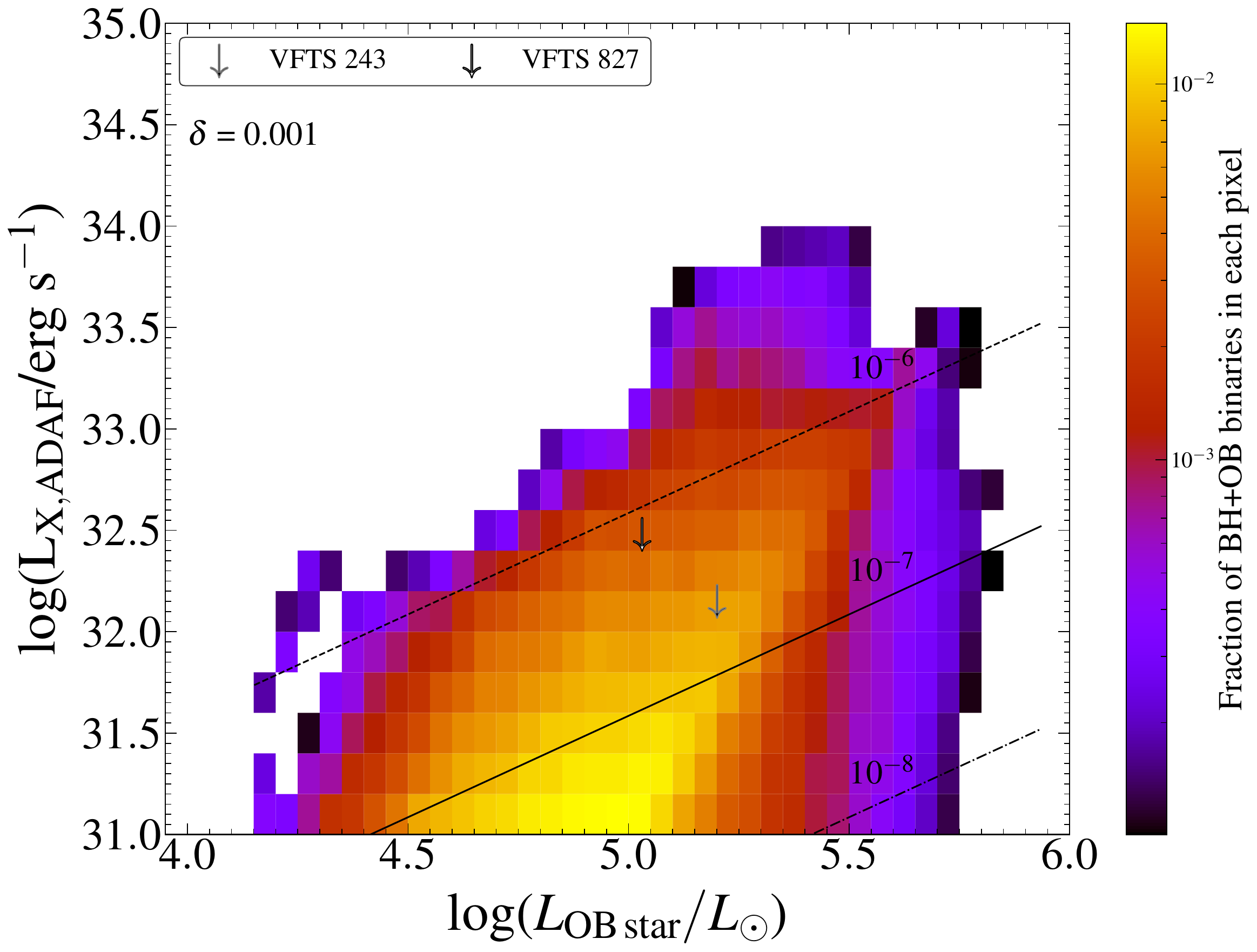}
    \includegraphics[width=0.48\hsize]{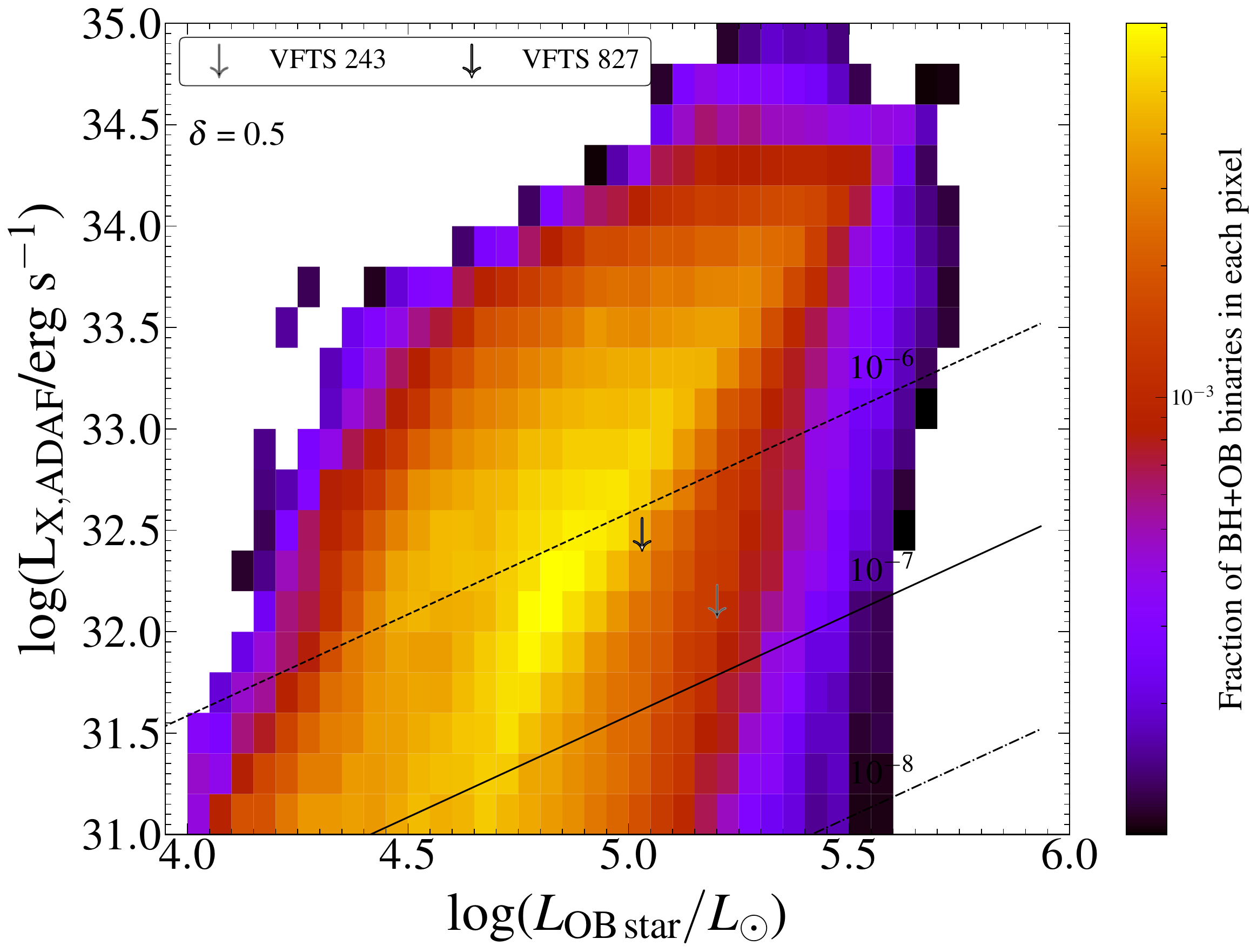}
    \caption{Distribution of X-ray luminosity produced from ADAFs around BHs without Keplerian accretion disks and the bolometric luminosity of the OB star companion, for two values of the viscous dissipation parameter $\delta$ = 0.001 (\textit{left panel}) and 0.5 (\textit{right panel}). The colour bar gives the predicted number of BH+OB star binaries in each pixel. The sum of the pixels in each panel equals unity. The solid black line shows the $L_{\rm X,w}$ = 10$^{-7}$ $L_{\rm OB}$ relation, with the dotted line and the dot-dashed line showing the $\pm$1\,dex range. The black and grey downward arrow denotes the upper limit to the X-ray luminosity from VFTS\,827 and VFTS\,243. }
    \label{figure_Lx_Lo}
\end{figure*}

\section{Discussion}
\label{section_discussion}

\subsection{Disentangling thermal from non-thermal X-ray emission}
\label{section_disentangling}

Figure\,\ref{figure_Lx_Lo} shows the distribution of X-ray luminosity 
from ADAF and OB star luminosity for the BH+O star binaries 
that do not form a Keplerian accretion disk. For inefficient viscous 
heating (left panel), the peak of the distribution lies below the canonical 
wind X-ray emission$L_{\rm X,w}$ = 10$^{-7}$ $L_{\rm bol}$ relation 
\citep[][solid black line]{Crowther2022}. For efficient viscous heating 
(right panel), the peak of the distribution is nearly an order of magnitude 
above the typical wind X-ray luminosity expected from the main sequence 
companion to the BH. We suggest that targeted X-ray observations on the 
B stars in the SB1 sample of \citet{Shenar2022} (see Table\,\ref{table_observations}) 
provide a suitable opportunity to investigate the presence of BHs, 
and study accretion physics around stellar mass BHs. 
We identify two B stars, VFTS\,186 and VFTS\,640, in the T-ReX catalogue 
that show $L_{\rm X,w}$/$L_{\rm bol}$ > 10$^{-6}$. However, we note that 
\citet{Evans2015} categorises them as single stars based on radial 
velocity measurements. We identify two more O stars with 
$L_{\rm X,w}$/$L_{\rm bol}$ > 10$^{-6.5}$ in the T-ReX catalogue of 
\citet[][see their Table A2, T-ReX label cc4651, c7552]{Crowther2022} 
showing an X-ray excess, above the canonical wind X-ray luminosity. 
The X-ray excess could be ascribed to accretion onto an orbiting BH, 
a possibility which deserves follow-up observations. Finally, we notice 
that the non-detection of X-rays from the black hole candidate VFTS 
243 sets an upper limit to the X-ray emission of 
$\sim 2\times 10^{-7}L_{\rm bol}$, marginally above the expected 
emission from the wind \citep{Shenar2022n}.

Even in systems where the X-ray luminosity due to accretion is lower 
than the X-ray intrinsic emission from the stellar wind, detecting 
the former is not beyond reach. Indeed, their spectral properties 
are fundamentally different \citep[e.g.,][]{Gierlinski1999}. In the 
radiatively inefficient regime, the X-ray luminosity from the low-density 
plasma near the BH is non-thermal (see section\,\ref{section_synchrotron}). 
For instance, synchrotron emission from relativistic electrons 
accelerated by magnetic reconnection would yield a power-law of 
spectral index of 0 \citep{ElMellah2022} to -0.7 \citep{Ponti2017}, 
depending on the efficiency of radiative cooling. In the case of a 
strong magnetic field, this emission could peak in hard X-rays, 
beyond the maxima from wind X-ray emission. This spectral disentangling 
between the thermal emission of the stellar wind and the non-thermal 
emission from the dilute accretion flow was used by \citet{Munar-Adrover2014} 
to measure an X-ray luminosity of from the X-ray faint system MWC 
656, later refined to $L_X\sim 4\times 10^{-8}L_{\rm bol}$ \citep{Ribo2017}. 
Even if the presence of the BH in this system has been challenged 
in a recent work \citep{Janssens2023}, it is a promising precedent 
which illustrates that non-thermal emission can be detected below 
the thermal wind X-ray luminosity level. 

\subsection{Be X-ray binaries and BH+OBe binaries}
\label{section_be_xray_binaries}

The typical X-ray luminosity of Be X-ray binaries is higher than 
that expected from advection-dominated accretion in the BH+OB 
star binaries. The predicted 
and the observed distribution of the outburst X-ray luminosity 
ranges from $10^{\rm 34} - 10^{\rm 39}$\,erg\,s$^{-1}$ 
(\citealp{Cheng2014} and \citealp[Fig. 4 of][]{Liu2024}). 
Hence, the X-ray luminosity of the majority of X-ray-faint 
BH+OB star binaries predicted in this work are not expected to overlap 
with the X-ray luminosity from the population of fairly brighter 
Be X-ray binaries or the BH+OB star binaries having a Keplerian accretion 
disk around the BH. We also note that all the known Be 
X-ray binaries are found to host a neutron star, not a BH 
\citep{Ziolkowski2014,Brown2018}.

We show in Fig.\,\ref{figure_vrot_distribution} that more than 
80\% of our BH+OB star systems will host an OB star that rotates 
at more than 80\% of its critical rotational velocity. As such, 
we expect the majority of the predicted X-ray-faint BHs to have an 
OBe companion. In absolute numbers, our models predict $\sim$109 
BH+OBe systems in the LMC, amongst the $\sim$4500 photometrically 
detected OBe stars in the LMC \citep{Schootemeijer2022}. BH+OBe 
systems have rarely been observed (with one candidate being AS386, 
\citealp{Khokhlov2018}). Our results provide an alternative way to 
search for these elusive BH+OBe systems. 

Smoothed Particle 
Hydrodynamics simulations done by \citet{Brown2018} show 
that for an eccentricity of 0, the average X-ray luminosity 
due to mass accretion onto the BH from the decretion disk of 
the OBe star in a BH+OBe system is around $10^{31}$\,erg\,$s^{-1}$ 
(see their figures 6 and 7). We also note that \citet{Brown2018} 
assume a radiative efficiency of 0.1 in their X-ray luminosity 
estimates (see their equation 4), which may overestimate their 
X-ray luminosity prediction. Moreover, \citet{Zhang2004} propose 
that the decretion disk can get significantly truncated in the 
presence of a BH. Based on the above studies, the X-ray luminosity, 
due to mass accretion onto the BH from the decretion disk around 
the OBe star of a BH+OBe star binary, may not be strong enough 
to overlap with our predictions for X-ray luminosity from ADAF 
around the BH.

\subsection{BH spin}

\citet{Qin2018,Fuller2019} showed that the black hole formed from the 
initially more massive star in a binary has a negligible spin parameter 
\citep[see also][]{Marchant2023}. Moreover, a BH needs to accumulate mass 
equal to its mass to significantly increase its natal spin parameter 
\citep{King1999,Wong2012}, which is not feasible from sub-Eddington 
mass accretion (Fig.\,\ref{figure_Macc_Mnet}) over a timescale of a 
few Myrs. This is reflected in our assumption that the spin of the BH 
during the BH+OB star phase is zero. \citet{Sen2021} showed that the 
formation of an accretion disk is favoured in the case of more rapidly 
spinning black holes, as the innermost circular orbit radius of the 
black hole decreases with increasing BH spin. Hence, the predicted 
number of X-ray-bright BH+OB star binaries will increase for spinning 
black holes.

Previous studies of X-ray-bright BH+OB star binaries such as Cyg\,X-1 
\citep{Gou2011,Miller-Jones2021,Zhao2021} and LMC\,X-1 \citep{Gou2009,Mudambi2020} 
have inferred that the black holes in these systems are maximally spinning 
($a_{\rm BH}$ > 0.90). On the other hand, recent simulations have found 
that deduced spins of wind-fed BH HMXBs may be model-dependent \citep{Belczynski2021,Zdziarski2023}. 
The X-ray spectral energy distribution can be statistically 
well-fitted with the assumption of a slowly spinning black hole and a 
Comptonised layer above the Keplerian accretion disk. Our population 
synthesis results, based on the accretion disk formation criteria of 
\citet{Sen2021}, show that $\sim10$ non-spinning BHs in BH+OB star 
binaries can form accretion disks around them in the LMC, relaxing 
the necessity for maximally spinning black holes to form wind-fed BH 
HMXBs. Furthermore, \citet{Batta2017} showed that a \textit{failed} 
supernova explosion with sufficient fallback accretion can produce 
rapidly spinning BHs with a spin parameter > 0.8, while the expected 
spin parameter from direct collapse is < 0.3. 

\subsection{Accretion disk formation criterion}
\label{section_accretion_disk_formation}

Our criterion with $f(\bar{s})=1$ overpredicts the number of observable 
wind-fed BH HMXBs in the LMC by an order of magnitude. For 
$f(\bar{s})=\sqrt{12}$, Eq.\,(\ref{eq:disk_formation}) predicts 
$\sim$1.5\% of the BH+OB binaries are expected to form an accretion 
disk. This leads to only $\sim$2 predicted X-ray-bright BH+OB star 
binaries in the LMC, consistent with one observed wind-fed BH HMXB, 
LMC X-1. Furthermore, for $f(\bar{s})=\sqrt{12}$, the number of 
X-ray-faint BHs are 35$\dots$51$\dots$80 for $\delta$ = 0.001, 0.1 
\& 0.5. 

Recent rapid 
binary population synthesis simulations by \citet{Romero2023} showed 
that an accretion disk formation criterion that solely relies on the 
Roche lobe filling factor \citep{Hirai2021} of the OB star companion 
may explain the scarcity of wind-fed BH HMXB in the Milky Way. However, 
their criterion was derived from a specific set of binary parameters, 
especially a mass ratio of 2 ($M_{\rm OB}/M_{\rm BH}$) and orbital 
periods up to $\sim20$\,d. At lower mass ratios, we find that the wind 
speed of an OB star companion to the BH may be low enough to form an 
accretion disk even at orbital periods of 10s of days 
(Fig.\,\ref{fig:disk_form}). 

If we additionally constrain our accretion disk formation criterion 
such that the OB companion must have a Roche lobe filling factor above 
0.85 (see Fig.\,\ref{fig:disk_form}, we find that our models predict 
$\sim$2.5 X-ray-bright BH+OB star systems in the LMC. However, we do 
not see any physical reason to discount the BH+OB star binaries at 
long orbital periods that may also be observable as wind-fed BH HMXBs. 
We also note that the Roche lobe filling factor of the O star in 
HD96670 is $>$0.85, but no bright X-ray emission ($>$10$^{35}$\,erg\,s$^{-1}$) 
has been detected. A further investigation into the accretion disk 
formation may be useful, but beyond the scope of this work. 

\section{Conclusion}
\label{section_conclusion}

Understanding the production of X-rays from BH+OB star binaries 
is essential to explaining the observed population of BH high-mass 
X-ray binaries \citep{Fornasini2023,Zhang2024}. In this work, we 
study the population of BH+OB star binaries that are not expected 
to have a Keplerian accretion disk around the BH \citep{Langer2020,Sen2021}. 
These binaries are expected to be the higher mass counterparts of 
the population of Be X-ray binaries \citep{Vinciguerra2020,Liu2024}, 
but without copious X-ray emission \citep{Sharma2007,Xie2012}. 
Moreover, these X-ray-faint BH+OB star binaries are the tentative 
progenitors of the binary BHs that can merge within Hubble 
time \citep{Belczynski2008,Romero2023} from the isolated binary 
evolution channel. 

We use $\sim$20\,000 detailed binary evolution models that include 
differential rotation, time-dependent tidal interactions and angular 
momentum transport during the mass transfer phases. Only $\sim$8\% 
of the total population of BH+OB star binaries in our grid are expected 
to form wind-fed BH HMXB systems like LMC\,X-1. Our models predict at 
least $\sim$28 BH+OB star binaries to be observable in the LMC with 
current X-ray telescopes like \textit{Chandra} and the upcoming 
SRG/eROSITA ($L_{X}$ = $10^{\rm 31}$-$10^{\rm 35}$\,erg\,s$^{-1}$). 

We show in Sect.\,\ref{section_radiative_efficiency} that viscous 
heating imparted to electrons can be high in the presence of magnetic 
fields of $\sim$30\,G, whether the flow is collisional \citep{Xie2012} 
or not \citep{Sharma2007}. For efficient viscous heating, our models 
predict that a significant fraction of BH+OB star binaries without 
accretion disks can produce X-ray luminosity between 
$10^{\rm 31}$-$10^{\rm 35}$\,erg\,s$^{-1}$ (Sect.\,\ref{section_numbers}). 
Up to $\sim$72 X-ray-faint BH+OB star binaries in the LMC may be observable 
due to high viscous heating efficiency in ADAFs around stellar mass 
black holes. 

Recent photometric and spectroscopic observations into the massive star 
content of the LMC provides an excellent testbed for our model predictions 
\citep{Evans2011,Evans2015,almeida2017,mahy2019a,mahy2019b}. Identifying 
a population of X-ray-faint BH+OB star binaries from recent observational 
samples of SB1 binaries \citep{Villasenor2021,Shenar2022,Mahy2022,Banyard2023} 
will provide crucial constraints on the contribution of the isolated binary 
evolution channel to the population of double BH mergers. 

Our study of faint X-ray emission from BH+OB star binaries makes the case 
for long-time exposure surveys in the X-ray band on tentative BH+OB star 
candidate systems without an X-ray-bright counterpart. Our method may also 
be applied to isolated BHs to produce complementary predictions in the X-ray 
band. This may be useful for future all-sky multi-wavelength surveys with 
high sensitivity to hunt for the large unidentified population of BHs in 
the Milky Way \citep{Scarcella2021}. 

\begin{acknowledgements}
      KS is funded by the National Science Center (NCN), Poland, under grant number OPUS 2021/41/B/ST9/00757. The authors thank Feng Yuan, Ying Qin and the referee Yong Shao for their valuable comments. The authors also thank Pablo Marchant for using his models. 
\end{acknowledgements}

\bibliographystyle{aa}
\bibliography{faint}

\begin{appendix}

\section{Predictions for $\delta$ = 0.1}

\begin{figure}[H]
    \centering
    \includegraphics[width=\hsize]{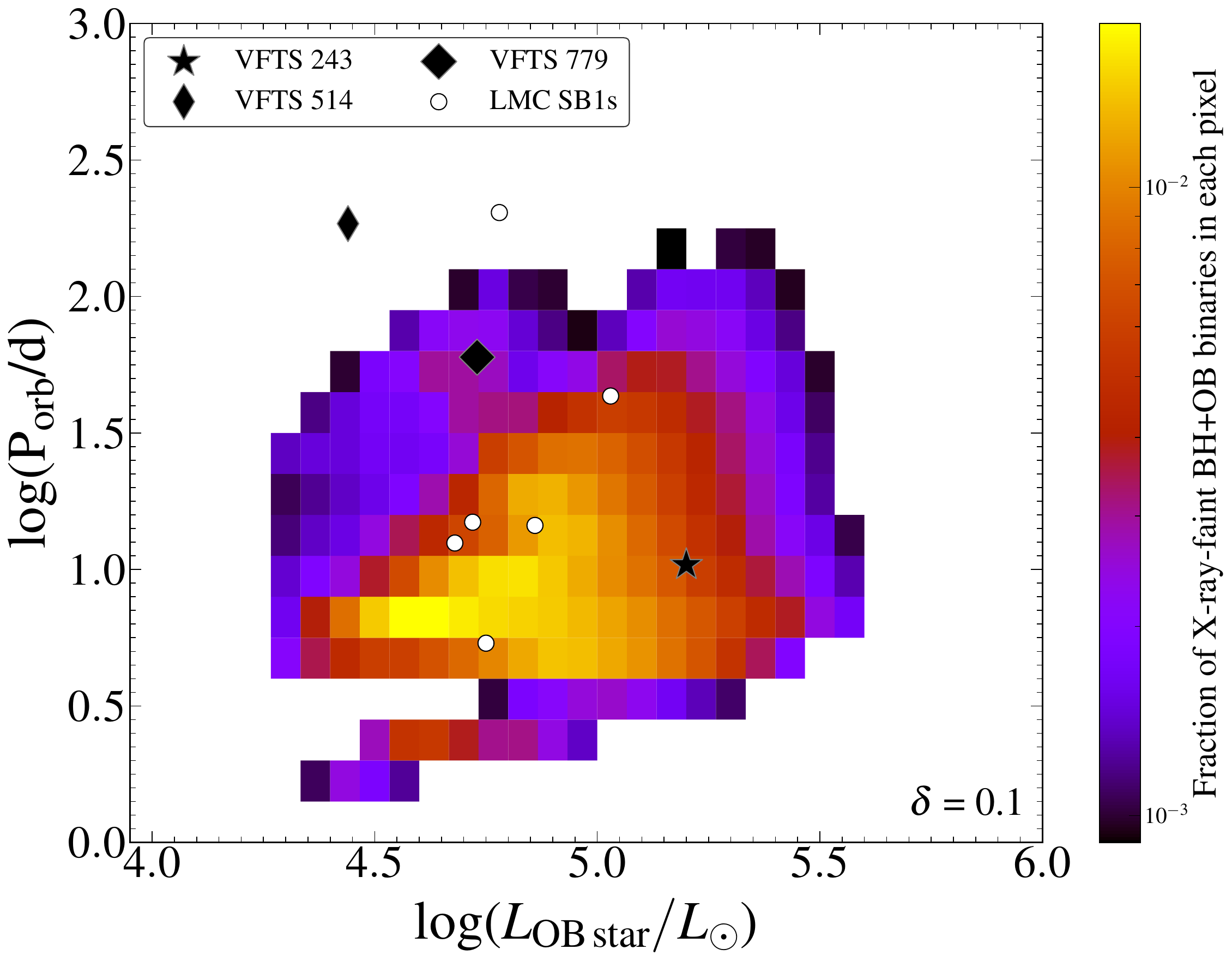}
    \caption{Same as Fig.\,\ref{figure_Porb_Lo} but for $\delta$ = 0.1. }
    \label{figure_Porb_Lo_appendix}
\end{figure}

\begin{figure}[H]
    \centering
    \includegraphics[width=\hsize]{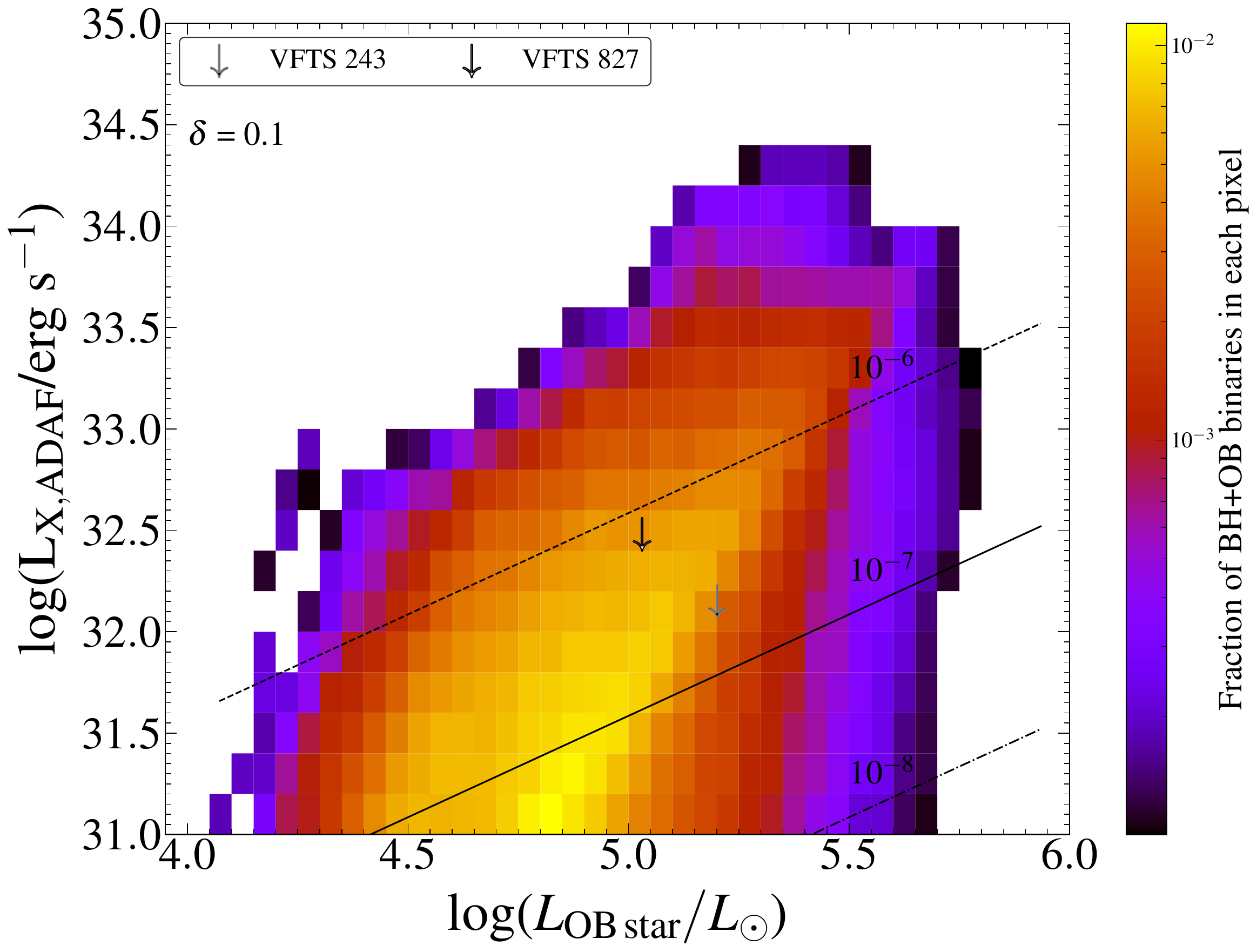}
    \caption{Same as Fig.\,\ref{figure_Lx_Lo} but for $\delta$ = 0.1. }
    \label{figure_Lx_Lo_appendix}
\end{figure}

\section{Observable parameter distribution for X-ray-bright systems}

\begin{figure}[H]
    \centering
    \includegraphics[width=\hsize]{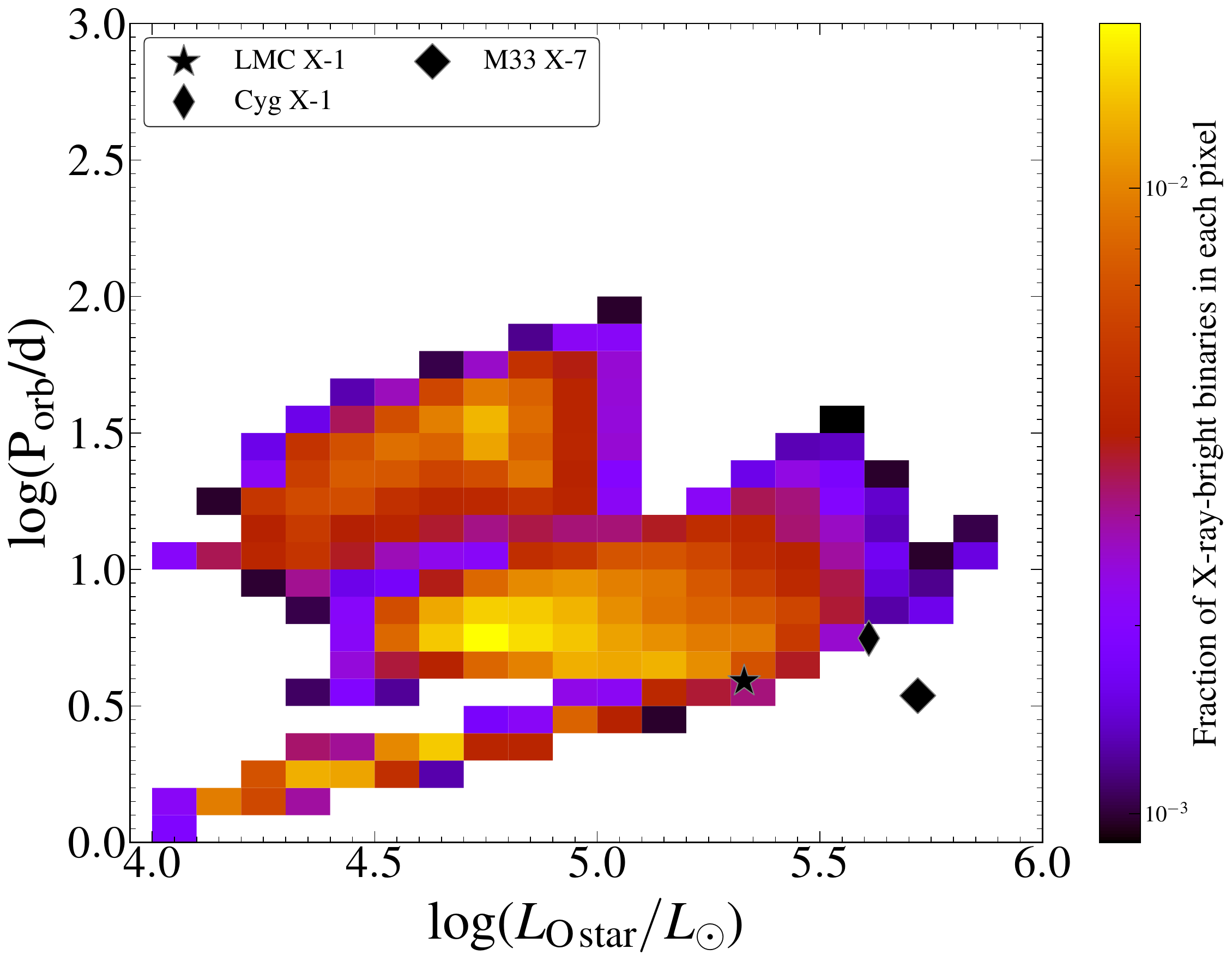}
    \caption{Same as Fig.\,\ref{figure_Porb_Lo} but showing the distribution of BH+OB star binaries that can produce X-ray luminosity above 10$^{35}$\,erg/s. The black markers show the position of the three observed wind-fed BH HMXBs. }
    \label{figure_Porb_Lo_X-ray-bright}
\end{figure}

\section{Rotational velocity distribution of the OB companions}

\begin{figure}[H]
    \centering
    \includegraphics[width=\hsize]{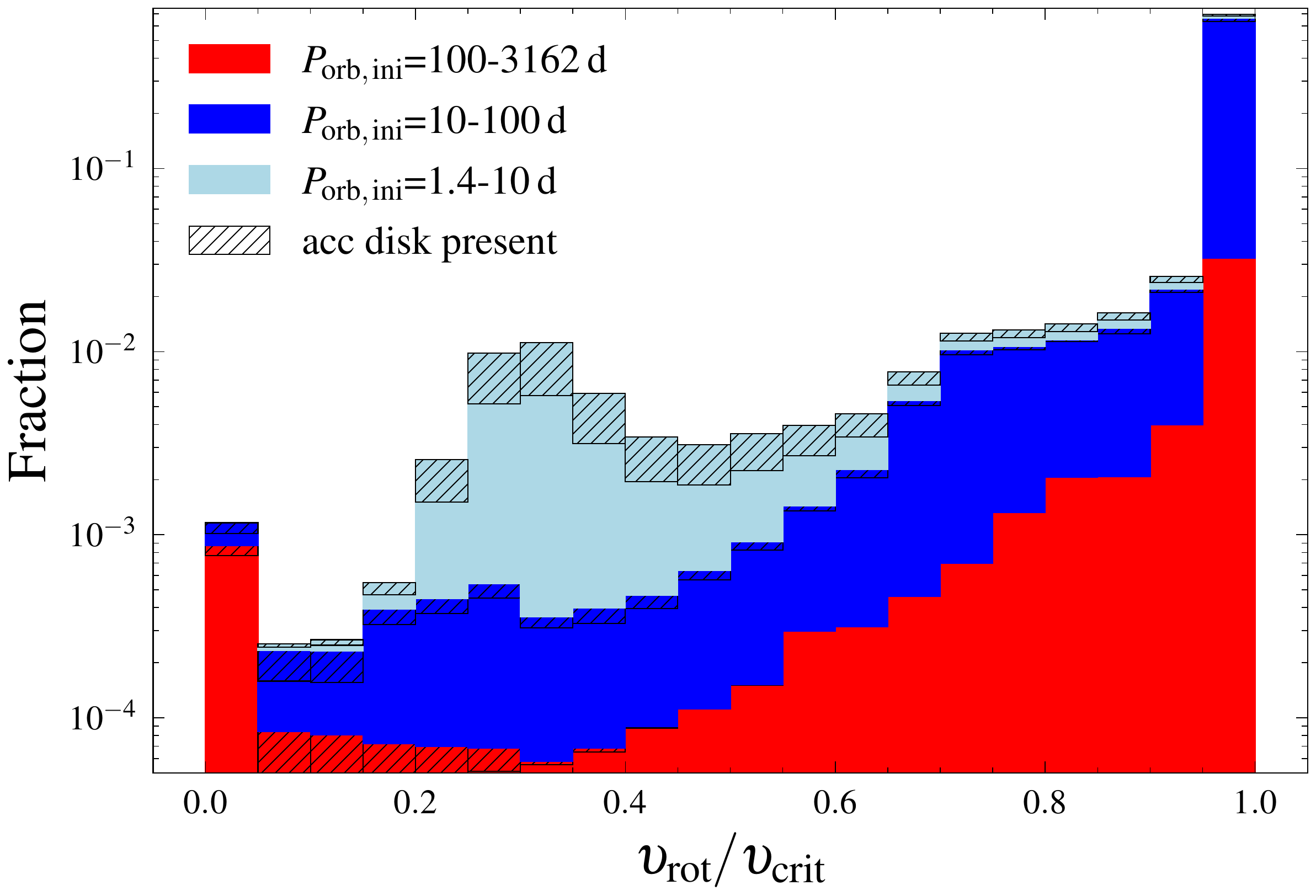}
    \caption{The ratio of the rotational velocity to the critical rotational velocity of the OB star companion during the BH+OB phase for all our binary models. The three colours denote the contributions from different initial orbital period ranges. The histograms are stacked on top of each other and the sum of ordinate values equals unity. }
    \label{figure_vrot_distribution}
\end{figure}

Figure\,\ref{figure_vrot_distribution} shows the distribution of the 
ratio of the rotational velocity to the critical velocity of the OB 
star companions to the black hole during the BH+OB phase. In our models, 
about $\sim$72.5\% and $\sim$80.8\% of the OB stars during the BH+OB 
phase are rotating above 95\% and 80\% of their critical rotational 
velocity, respectively. As such, they may be observable as OBe stars 
\citep[e.g.][]{Shao2014,Schootemeijer2022}.

\end{appendix}

\end{document}